\documentclass[aps]{revtex4}
\usepackage{amsthm}
\usepackage{amsmath}
\usepackage{epsfig}
\def\abs#1{{\left\vert #1 \right\vert}}
\def\norm#1{{\left\Vert #1 \right\Vert}}
\def\trace{\mathrm{Tr}}
\def\nchoosek#1#2{\genfrac{(}{)}{0pt}{}{#1}{#2}}
\def\ket#1{\left\vert #1 
\right\rangle}
\def\bkm#1#2{\left\langle #1 \vert #2 \right\rangle}
\def\tensorm{\otimes}
\def\id{{\mathcal{I}}}
\def\bra#1{\left\langle #1 \right\vert}
\def\ensavg#1{\left\langle #1 \right\rangle}
\def\v#1{{\bf #1}}
\def\refsec#1{Sec.\ \ref{Section::#1}}

\def\refeqn#1{Eq.\ (\ref{Equation::#1})}

\def\refeqs#1#2{Eqs.\ (\ref{Equation::#1}) and (\ref{Equation::#2})}
\def\refeqto#1#2{Eqs.\ (\ref{Equation::#1}--\ref{Equation::#2})}

\newtheorem{theorem}{Theorem}

\begin{document}

\title{Non-Perturbative Bounds on Hyperfine-Induced
Electron Spin Coherence Times}

\author{Neil Shenvi}
\author{Rogerio de Sousa}
\author{K. Birgitta Whaley}
\affiliation{Department of Chemistry and the Kenneth
S. Pitzer Center for Theoretical Chemistry, University of California,
Berkeley, Berkeley, CA 94720}

\date{\today}

\begin{abstract}
We address the decoherence of a localized electron spin in an 
external magnetic field due to the hyperfine interaction with a lattice 
of nuclear spins.  Using a completely 
non-perturbative method, rigorous bounds on the $T_1$ and $T_2$ coherence 
times for the electron spin are provided.  It is shown that for magnetic 
fields $B$ greater than some 
critical field $B_c$ ($B_c \approx .001 - 2\, \mathrm{Tesla}$ for the 
systems studied here), the $z$-polarization of the electron spin cannot 
relax, and hence $T_1$ is infinite.  However, even at high fields 
dephasing can still occur.  We provide a lower bound on the $T_2$ 
coherence time that explicitly takes into 
account the effects of a spin echo pulse sequence.
\end{abstract}  


\maketitle

\section{Introduction}
Isolated electron spin systems are of great interest because of their 
potential application as coherent quantum memory in novel 
solid state devices\cite{Wolf:01,Zutic:04RMP}.  Several quantum 
computation proposals make use of electron spins, either as the qubits 
themselves or as intrinsic components of one- and two-qubit gates 
\cite{Loss:98A, Burkard:99B, Kane:98, Skinner:03L}.  Furthermore, 
experimental efforts are just 
now succeeding at measuring single electron spin relaxation times 
\cite{Tyryshkin:03B, Abe:04X, Elzerman:04}.  
Unfortunately, the exponential 
size of the Hilbert space of these spin systems and the variety of 
channels for decoherence makes an 
exact analysis of spin dynamics virtually impossible (without the use of a 
quantum 
computer)\cite{Khaetskii:03B,deSousa:03B1,Mozyrsky:02B,Semenov:03L}.  
As a result, it is often necessary to resort to 
approximations or perturbative treatments to address the timescale of 
electron spin decoherence, even using a simplified Hamiltonian.  
Because approximations are inevitable, it is important to know which of 
these can be 
rigorously justified.  In particular, it is necessary to know which 
terms in 
a very complex system Hamiltonian can be ignored because they do not 
contribute to decoherence on the experimentally relevant timescale.
In this paper, we consider electron spin decoherence induced by the 
nuclear spins.  In particular, we focus on the relevance of the $S_+ 
I_- + S_- I_+$ term in the Hamiltonian (see \refeqn{HFull}), which 
comes from the contact hyperfine coupling of the electron and 
its surrounding nuclei.

Previously, the effect of the hyperfine interaction on decoherence has been 
studied in special cases or using various approximation 
methods.  Recently, Erlingsson et. al used a semi-classical 
approximation to study electron spin correlation functions at low 
fields ($B \ll B_c$, where $B_c$ is on the order of the Overhauser 
field, see below) \cite{Erlingsson:04}.  They found that the
correlation functions were characterized by complex dynamics which
did not decay, even at long times.  Work by Khaetskii et al. has 
characterized the timescale 
of the logitudinal spin correlation for an electron coupled to 
surrounding nuclei via the hyperfine interaction
\cite{Khaetskii:03B,Khaetskii:03L}.  
In  refs. \cite{Khaetskii:03B,Khaetskii:03L}, exact solutions for the 
case of 
completely polarized nuclei and for the case of 
homogeneous coupling, as well as perturbative solutions for the 
general, inhomogeneous situation were presented.  From these solutions, 
the behavior of the general unpolarized, inhomogeneous case was inferred.
Their results indicate that the logitudinal spin correlation time is on the 
order of $\hbar N / A$ for a single quantum dot and on the order of 
$\hbar \sqrt{N}/A$ for an ensemble of dots.
Gate imperfections due to 
the hyperfine coupling were also treated using perturbation theory in 
\cite{Saikin:03B}.  However, it was noted in \cite{Khaetskii:03L} that the 
perturbative solution would diverge to second-order and was therefore 
potentially unreliable.  More recently, Coish et al. have treated the 
behavior of this system using a generalized master 
equation, again confirming a power-law decay for short
timescales on the order of $\hbar N / A$ \cite{Coish:04X}.  
In general, though, all 
treatments based on a time-dependent perturbation expansion will be valid 
only at times smaller than some maximum timescale specified by the order 
of the expansion\footnote{For instance, in \cite{Coish:04X}, the continuum 
hypothesis is only strictly valid for timescales $t \ll \hbar N^{3/2} / A$ 
[W.A. Coish, Private communication]}.  
Although such treatments are useful for examining
short-timescale and small-magnitude fluctuations in electron polarization, 
in this paper we are interested in the long-timescale dynamics of the 
electron spin which leads to $1/e$ decay of 
the longitudinal or in-plane 
polarization of the electron.  Note that we are not claiming any 
particular 
functional form for the decay of coherence, exponential or otherwise.  
Rather, we are evaluating the timescale on which the longitudinal and 
in-plane magnetization decays to a constant fraction (in this case $1/e$) 
of its initial value.  Moreover, we focus on the high magnetic 
field regime which is relevant for the recent experimental 
developments in single spin read-out ($B = 8-14\,\mathrm{Tesla}$ in 
ref. \cite{Elzerman:04}).

In ref. \cite{deSousa:03B1} it was assumed that at sufficiently 
high magnetic fields the electron-nuclei flip-flop term ($S_+ I_- + 
S_- I_+$ in \refeqn{HFull}) which is due to the hyperfine interaction 
can be neglected as a direct source for electron spin decoherence.  In 
this paper, we would like to 1) justify this 
assumption and 2) determine how strong a magnetic field must be applied 
to safely neglect this term.
Rather than resort to approximation to estimate the numerical value of $T_1$ 
and $T_2$, this paper will instead place rigorous, non-perturbative lower 
bounds on 
the spin relaxation time of the electron in the high-field regime ($B > 
B_c$).  Furthermore, we will 
explicitly take into account the application of the spin echo sequence 
which is often used experimentally to remove inhomogeneous broadening.
To determine the importance of the electron-nuclear flip-flop term, we 
use 
the Hamiltonian presented in \cite{Khaetskii:03B,Khaetskii:03L,Coish:04X} which 
contains only the contact hyperfine term.  In doing so, we can 
effectively isolate the decoherence contribution that derives exclusively 
from the hyperfine 
interaction.  When the timescale of this 
hyperfine-induced decoherence is significantly greater than the timescale 
of decoherence due to other mechanisms (such as spectral 
diffusion induced by inter-nuclear dipolar coupling\cite{deSousa:03B1}) 
then it seems reasonable to ignore the direct 
effects of hyperfine coupling as a channel for electron spin decoherence.  

It is important to define what we mean by ``electron spin 
decoherence''.  Typically, electron spins are assumed to undergo
simple exponential decay governed by the Bloch equations.  Thus
the coherence times $T_1$ and $T_2$ can be defined as the timescales
of this exponential decay.  However, when decay is non-exponential, as in 
this system (see \cite{Khaetskii:03B}), there is some flexibility in the
definition of the coherence times.  Electron spin dynamics includes both 
short-timescale fluctuations (which are generally small in magnitude 
\cite{Coish:04X}) and long-timescale (i.e., $1/e$) decay; in this paper, 
we take ``electron spin decoherence'' to mean the latter.  In other words, 
the coherence time of an electron is defined as the time it takes for the 
longitudinal or transverse polarization of an initially polarized 
electron spin to decay to a specific fraction, e.g. $1/e$, of 
its initial value.  

The decay of the in-plane magnetization is understood to occur in the context of a spin echo 
experiment.  Spin echo experiments are designed to remove the effects of 
inhomogeneous broadening due to variations in the local field experienced by an 
ensemble of electron-nuclei systems (i.e. multiple independent systems).  
When applied to a single electron spin, a spin echo can be understood as removing the 
inhomogeneous broadening resulting from an ensemble of initial states of the single 
electron-nuclei system.  The remaining decay of the in-plane magnetization comes from fluctuating 
Overhauser fields, which cannot in general be removed by the spin echo (but see 
\cite{Shenvi:04X2} for numerical evidence that a 
substantial part of hyperfine-induced fluctuations may, in fact, be 
reversible).  Because only 
\emph{bounds} on relaxation behavior are desired, our analysis of this decay can be completely 
rigorous, despite the complexity of the problem and the size of the Hilbert 
space.  Our results are applicable to localized electron spins in 
semiconductors such as quantum dots and donor impurities.

This paper is organized as follows: Sec. II outlines the general 
structure of the problem and presents the 
results of our analysis.  The derivation of the theorems used in Sec. II can be found in the Appendix on 
EPAPS\footnote{See 
EPAPS Document No. [number will be inserted by publisher] for the derivations of the theorems
presented in Sec. II.  A direct link to this document may be 
found in the online article's HTML reference section. The document may also be 
reached via the EPAPS homepage (http://www.aip.org/pubservs/epaps.html) or from 
ftp.aip.org in the directory /epaps/. See the EPAPS homepage for more information.}.  Applications to electron spins in 
semiconductors of relevance 
to solid state quantum information processing are made in Sec. III.  Discussion and conclusions follow in Sec. IV.

\section{Results}
Let $\v{S}$ be the electron spin precessing in a magnetic field $B$ 
applied in the $z$ direction.  The electron spin is coupled to $N$ 
nuclear spins $\v{I}_1, \v{I}_2, \ldots, \v{I}_N$ via the hyperfine 
interaction, characterized by the coupling constants $A_1, A_2, \ldots 
A_N$.  Due to the hyperfine coupling, the nuclei induce an effective 
Overhauser field on the electron with magnitude $\sum_j{A_j \v{I}_j}$.
Here we consider the Hamiltonian
\begin{eqnarray} \label{Equation::HFull}
H &=& \gamma_S B S_z + \gamma_I B \sum_j{I_{jz}} +\sum_j{A_j \left(S_z 
I_{jz} + \frac{1}{2}\left(S_+ I_{j-} + S_- I_{j+}\right)\right)},
\end{eqnarray}
where $\gamma_S$ and $\gamma_I$ are the gyromagnetic ratios 
of the electron and nuclei, respectively ($\gamma_S \approx 10^3 
\gamma_I$).
\refeqn{HFull} can be written as
\begin{eqnarray} \label{Equation::HFlipFlop}
H &=& H_0 + V, \\
H_0 &=& \gamma_S B S_z + \gamma_I B \sum_j{I_{jz}} + 
\sum_j{A_j S_z I_{jz}}, \\
V &=& \sum_j{\frac{1}{2}A_j\left(S_+ I_{j-}+S_- I_{j+}\right)}.
\end{eqnarray}
This Hamiltonian conserves the total spin angular momentum $J_z = S_z + 
\sum_j{I_{jz}}$; hence, the total number of ``down'' spins is 
conserved.  For convenience, we will label each 
of these blocks by the quantum number $L$, where $L$ is the total number 
of 
``down'' spins (i.e. $J_z = N-2L$, $L = (N-J)/2$).  Then the
Hamiltonian can be block diagonalized based on the $J_z$ operator 
\cite{Coish:04X}.  
There will be $N+2$ blocks, one for each possible value of $L$, and the 
$L^{th}$ block will be an $\nchoosek{N+1}{L}\times\nchoosek{N+1}{L}$ 
matrix.
For the remainder of the paper, it will be assumed that all analysis 
applies to some subspace labeled by a particular value of 
$L$.  

The unperturbed Hamiltonian, $H_0$, commutes 
with the $S_z$ operator.  Thus, the eigenstates of $H_0$ can be labeled 
by the $z$-polarization of the electron.  
On the other hand, the full Hamiltonian $H$ does not commute with $S_z$ 
because $V$ connects electron spin-up states to electron spin-down states, 
and vice-versa.  At high fields greater than some critical field 
$B > B_c$, the Zeeman energy of the electron will dominate the energy 
of the eigenstates, and the electron spin-up and electron spin-down 
manifolds will be well separated in energy.  Thus, if the perturbation is 
in some sense smaller in magnitude than this energy gap, the effects of 
the perturbation will be small and $S_z$ will still be an approximately 
good label for eigenstates of $H$ (see Figure 1).

Our analysis in Sec. II.A demonstrates that if the magnetic field $B$ is 
greater than the critical field $B_c$, then the electron spin magnetization 
$\ensavg{S_z}$ never changes appreciably.  Because the eigenstates of $H$ are nearly eigenstates 
of $S_z$ at high fields, the $z$-polarization of the electron does not undergo 
substantial (i.e. $1/e$) decay (this result matches the perturbative 
result found in \cite{Khaetskii:03L}).
Furthermore, based on these arguments, a field-dependent bound is placed 
on the maximum fluctuation of $S_z$ from its initial value: the greater 
the field, the smaller the maximum possible fluctuation.
Again, it should be emphasized that there may be small magnitude fluctuations in 
$\ensavg{S_z(t)}$, which occur on a finite timescale.  However, because our definition 
of $T_1$ (the timescale for $1/e$ decay of $\ensavg{S_z}$) quantifies the large-scale 
coherence properties of the electron spin, the absence of large-scale decay 
renders $T_1$ effectively infinite.  Although $T_1$ is infinite for large magnetic fields, 
dephasing processes can still occur provided that they conserve the $z$-polarization of the 
electron spin.  Although this statement seems like a contradiction given that the 
perturbation, $V$, flips the electron spin, virtual 
transitions involving $V^2$ can nevertheless flip pairs of nuclear spins 
without affecting 
the electron spin.  For this reason, it 
is also important to place a lower bound on $T_2$.

Derivation of the lower bound on $T_2$ in Sec. II.B follows a different 
methodology than the bound on 
$T_1$.  Because $T_2$ is defined as the timescale of the decay of the 
spin echo envelope \cite{Hahn:50R}, a lower bound on the magnitude of 
$\ensavg{S_+}$ after a pulse echo sequence implies a lower bound on $T_2$.  
Specifically, the standard definition of $T_2$ is the minimal time, $t$ 
such that,
\begin{equation}
\abs{\ensavg{S_+(2t)}} = \frac{1}{2e},
\end{equation}
where a $\pi$ pulse is applied at time $t$ (we have dropped a factor of 
$\hbar$ for convenience).
It is assumed that for an infinite system, there will be no 
large-scale recurrence in $S_+(2t)$, but only dissipative behavior. 
Small magnitude recurrence, due for instance to echo modulation 
\cite{Rowan:65R}, 
may still exist, but large-scale recurrence on the order of $1/e$ is assumed 
to be absent.  Even if this large-scale recurrence does exist, the 
lower bound 
on $T_2$ is still valid.  In particular, if it can be shown that for some 
value of $t'$, $\abs{\ensavg{S_+(2t')}} \geq 1/2e$, it follows that $T_2 
\geq 
t'$.  

This bound on $\abs{\ensavg{S_+(2t')}}$ can be proven using a variational 
argument.  Every trial state $\ket{\phi}$ induces a probability 
distribution over the eigenenergies of $H$.  The mean of this energy 
distribution is given by the expectation value
\begin{equation}
\ensavg{E} = \bra{\phi}H\ket{\phi}.
\end{equation}
Likewise, the variance of this energy distribution is
\begin{equation}
\ensavg{(\Delta E)^2} = \bra{\phi}H^2\ket{\phi}-\bra{\phi}H\ket{\phi}^2.
\end{equation}
The variance of the energy will be equal to zero if and only if 
$\ket{\phi}$ is an eigenstate of $H$ (or a superposition of degenerate 
eigenstates).  For all other states which are superpositions of 
non-degenerate eigenstates, 
the variance will be greater than zero.  The essential feature of the 
theorems 
given in Sec. II.B is that the dynamics undergone by any state $\ket{\phi}$ 
are limited in timescale by the variance in energy of that state.  Using 
this idea, it can be shown that the dephasing processes of the system 
have a timescale governed by the energy
\begin{equation}
W_{max} = \sum_{j}{\frac{\hbar^4 A_j^2}{4(\hbar \gamma_S 
B)}}.
\end{equation}
Conceptually, our bound on $T_2$ makes mathematically rigorous the idea 
that the coherence time of a state is inversely proportional to its 
linewidth.

Before proceeding to the proof of these results, two important points must 
be made.  
First, this method constructs only a rigorous \emph{lower bound} on $T_2$; it 
does not predict the actual value of $T_2$ nor does it claim to be a 
\emph{tight} lower bound on $T_2$.  In fact, numerical work suggests that
the spin echo pulse may actually remove nearly all hyperfine-induced 
decoherence, leading to only small magnitude decay of $S_+(t)$ under 
the spin echo sequence for all times, $t$
\cite{Shenvi:04X2}.  However, this evidence is numerical and
has been confirmed only for small numbers of spins ($N \sim 10$).  In this
paper we are interested in what rigorous statements can be made regarding
the $T_2$ time for a large system ($N \sim 10^4 - 10^6$).

Second, it is the special structure of this system at high $B$ fields that allows a 
\emph{useful} bound to be constructed.  Though this method is general 
and can be used to construct valid bounds on $T_2$ for other systems, 
these bounds may prove to be useless in practice (for instance, the 
rigorous bound that $T_2 > 10^{-30} s$ is valid, but is a practically 
useless 
bound).  Fortunately, we find that our method provides practically useful 
lower bound values for real systems of relevance to quantum information 
processing (see \refsec{Application}).

\subsection{Bound on $T_1$}
The proof of the bound on $T_1$ will be given first and will be organized 
as follows.  First, we will show that the eigenstates 
of $H$ can be divided into two subspaces, the $+$ subspace and the 
$-$ subspace (Theorem 1).  Eigenstates contained in the $+$ subspace 
contain large contributions from electron spin-up states, and eigenstates 
in the $-$ subspace contain large contributions from electron spin-down 
states.  In other words, $S_z$ is an approximate quantum number for 
the true eigenstates of $H$.  Second, we will show that this fact places 
bounds on the eigenvalues of $H$ (Theorem 2).  Next, we will show that 
every eigenstate of $H_0$ is almost completely contained within either the 
$+$ or $-$ subspace of $H$ (Theorem 3).  To complete the bound on $T_1$, 
we will show that these previous statements imply a lower bound on the 
decay of $S_z$ from its initial state (Theorem 4).  

For future convenience, we define
\begin{eqnarray}
A[k] &=& \sum_{j \in k \, {\rm{largest}}}{A_j}\\
A_2[k] &=& \sum_{j \in k \, \rm{largest}}{A^2_j}\\
A_{max} &=& \sum_{j}{A_j} \\
A_{2,max} &=& \sum_{j}{A_j^2}.
\end{eqnarray}
In other words, $A[k]$ is the sum of the $k$ largest hyperfine coupling 
constants.  $A_{max}$ is the sum of all $N$ hyperfine coupling 
constants and therefore represents the maximum magnitude of the 
Overhauser field.
The matrix $H_0$ commutes with the $S_z$ and 
$I_{jz}$ operators; hence 
eigenstates of $H_0$ can be labeled by the $z$-polarization of the 
electron spin and the $z$-polarization of the nuclei.  
Let $\ket{\Uparrow}$ and $\ket{\Downarrow}$ represent the $+z$ and $-z$ 
polarized states of the electron spin, respectively, and let $\v{z}$ be an 
$N$-bit string of $+1$s and $-1$s representing the $z$-polarization of the 
$N$ nuclei.  The eigenstates of $H_0$ are of the form 
$\ket{\Uparrow,\v{z}}$ or $\ket{\Downarrow,\v{z'}}$ (it should be 
remembered that $\v{z}$ will 
contain $L$ down spins and $\v{z'}$ will contain $L-1$ down spins, since 
the total number of down spins, including the electron spin, is 
conserved).  The eigenvalue $E_{0,\v{z}}^\Uparrow$ corresponding 
to the eigenstate $\ket{\Uparrow,\v{z}}$ is
\begin{equation} \label{Equation::E0Plus}
E_{0,\v{z}}^\Uparrow = +\frac{1}{2}\hbar\gamma_S B + 
\frac{1}{2}\left(N-2L\right)\hbar\gamma_I B + \frac{1}{4}\hbar^2\sum_j{A_j 
z_j}.
\end{equation}
Similarly, the eigenvalue $E_{0,\v{z'}}^\Downarrow$ corresponding to 
the eigenstate $\ket{\Downarrow,\v{z'}}$ is,
\begin{equation} \label{Equation::E0Minus}
E_{0,\v{z'}}^\Downarrow = -\frac{1}{2}\hbar\gamma_S B + 
\frac{1}{2}\left(N-2L+2\right)\hbar\gamma_I B - 
\frac{1}{4}\hbar^2\sum_j{A_j z'_j}.
\end{equation}
From \refeqs{E0Plus}{E0Minus} it is clear that the eigenvalues of $H_0$ 
fall within certain ranges.  In particular, we can write
\begin{equation}
\begin{array}{rcccl}
E_{0,min}^\Uparrow &\leq& E_{0,\v{z}}^\Uparrow &\leq& E_{0,max}^\Uparrow \\
E_{0,min}^\Downarrow &\leq& E_{0,\v{z'}}^\Downarrow &\leq& E_{0,max}^\Downarrow,
\end{array}
\end{equation}
where we have defined
\begin{eqnarray} 
\label{Equation::E0Bound1}
E_{0,min}^\Uparrow &=& +\frac{1}{2}\hbar\gamma_S B + 
\frac{1}{2}\left(N-2L\right)\hbar\gamma_I B - 
\frac{1}{4}\hbar^2\left(2 A[L] - A_{max}\right) \\  
\label{Equation::E0Bound2}
E_{0,max}^\Uparrow &=& +\frac{1}{2}\hbar\gamma_S B + 
\frac{1}{2}\left(N-2L\right)\hbar\gamma_I B + 
\frac{1}{4}\hbar^2\left(2A[N-L] - A_{max}\right) \\
\label{Equation::E0Bound3}
E_{0,min}^\Downarrow &=& -\frac{1}{2}\hbar\gamma_S B + 
\frac{1}{2}\left(N-2L+2\right)\hbar\gamma_I B - 
\frac{1}{4}\hbar^2\left(2 A[N-L+1] - A_{max}\right) \\
\label{Equation::E0Bound4}
E_{0,max}^\Downarrow &=& -\frac{1}{2}\hbar\gamma_S B + 
\frac{1}{2}\left(N-2L+2\right)\hbar\gamma_I B + 
\frac{1}{4}\hbar^2\left(2A[L-1] - A_{max}\right).
\end{eqnarray}
It is critical to all proofs that follow that there is some energy gap 
$\Delta E$ between the electron spin-up and spin-down subspaces (see 
Figure 1).  In 
other words, the lowest energy electron spin-up state must be greater
in energy than the highest energy electron-spin down state.
Using \refeqs{E0Bound1}{E0Bound4}, it can be demonstrated that an energy 
gap exists if
\begin{equation} \label{Equation::ECondition}
\hbar (\gamma_S - \gamma_I) B \geq \frac{1}{2}\hbar^2 
\left(A[N-L]+A[N-L+1]-A_{max}\right).
\end{equation}
An additional condition that emerges from Theorem 2 sets the critical 
field slightly higher than \refeqn{ECondition} would imply.  
Furthermore, we would like the definition of $B_c$ to be independent of 
$L$.  Therefore, for the remainder of the paper, we will assume 
that we 
are in the high-field regime defined by
\begin{equation} 
B > B_c,
\end{equation}
where
\begin{equation} \label{Equation::BCondition}
B_c = \frac{\hbar A_{max}}{\gamma_S - \gamma_I}.
\end{equation}
Typical values of this critical field for various systems are given in 
Table 1.

Using these facts, we now show that the eigenstates of $H$ are 
near-eigenstates of $S_z$.  We define the projection operators 
$\Pi_\Uparrow$ and $\Pi_\Downarrow$ such that
\begin{eqnarray}
\label{Equation::PiUp}
\Pi_\Uparrow &=& \ket{\Uparrow}\bra{\Uparrow}\tensorm\id_I \\
\label{Equation::PiDown}
\Pi_\Downarrow &=& \ket{\Downarrow}\bra{\Downarrow}\tensorm\id_I,
\end{eqnarray}
where $\id_I$ is the identity operator on the nuclear spins.  Then the 
following theorem can be proven:

\begin{theorem}\label{Theorem::1}
The eigenstates of $H$ can be divided into two subspaces, 
which we will label $+$ and $-$.  An eigenstate $\ket{\psi^+}$ in the $+$ 
subspace obeys the conditions
\begin{eqnarray}
\bra{\psi^+}\Pi_\Uparrow\ket{\psi^+} &\geq& \xi_1 \\
\bra{\psi^+}\Pi_\Downarrow\ket{\psi^+} &\leq& \xi_2,
\end{eqnarray}
and an eigenstate $\ket{\psi^-}$ in the $-$ subspace obeys the conditions
\begin{eqnarray}
\bra{\psi^-}\Pi_\Downarrow\ket{\psi^-} &\geq& \xi_1\\
\bra{\psi^-}\Pi_\Uparrow\ket{\psi^-} &\leq& \xi_2,
\end{eqnarray}
where
\begin{equation} \label{Equation::Xi12Def}
\xi_{1,2} = \frac{1}{2}\pm\frac{1}{2}\sqrt{1-\frac{4 
V_{max}^2}{\left(E_{0,min}^\Uparrow 
- E_{0,max}^\Downarrow\right)^2}},
\end{equation}
such that $1\to +$, $2 \to -$, and
\begin{equation}
V_{max} = 
\frac{1}{2}\hbar^2 \sqrt{A[L] A[N+1-L]}.
\end{equation}
\end{theorem}

It is useful to note that
\begin{equation}
V_{max} \leq \frac{1}{2}\hbar^2 A_{max}
\end{equation}
for all L.  Above the critical field given in
\refeqn{BCondition}, we find that $\xi_1 > 1/2$ and $\xi_2 < 1/2$.  
Hence, we can write the eigenstates 
of $H$ as either $\ket{\psi_k^+}$ or $\ket{\psi_k^-}$, where the $+$ 
eigenstates are those which contain contributions primarily from electron 
spin-up states ($\ket{\Uparrow}$), and the $-$ eigenstates contain 
contributions primarily from the electron spin down states 
($\ket{\Downarrow}$),
\begin{eqnarray}
\bra{\psi_k^+}\Pi_\Uparrow\ket{\psi_k^+} &>& 1/2 \\
\bra{\psi_k^-}\Pi_\Downarrow\ket{\psi_k^-} &>& 1/2.
\end{eqnarray}
To summarize, we have shown that at high fields, states can still be 
approximately labeled according to their electron spin polarization.

We now proceed with the second step of the proof.

\begin{theorem}\label{Theorem::2}
The eigenvalue $E_k^+$ corresponding to eigenstate $\ket{\psi_k^+}$ 
and the eigenvalue $E_k^-$ corresponding to eigenstate $\ket{\psi_k^-}$ 
obey the following conditions
\begin{equation} \label{Equation::EPlusBound}
E_{min}^+ \leq E_k^+ \leq E_{max}^+
\end{equation}
and
\begin{equation} \label{Equation::EMinusBound}
E_{min}^- \leq E_k^- \leq E_{max}^-
\end{equation}
where
\begin{eqnarray}
E_{min}^+ &=& E_{0,\min}^\Uparrow - V_{max} \\
E_{max}^+ &=& E_{0,\max}^\Uparrow + V_{max} \\
E_{min}^- &=& E_{0,\min}^\Downarrow - V_{max} \\
E_{max}^- &=& E_{0,\max}^\Downarrow + V_{max}.
\end{eqnarray}
\end{theorem}
Theorem 2 demonstrates that the eigenvalues of the $+$ and $-$ eigenstates 
of $H$ fall in approximately the same ranges as the eigenvalues of the 
$\Uparrow$ and $\Downarrow$ eigenstates of $H_0$ (c.f. 
\refeqto{E0Bound1}{E0Bound4}).

The third step in the proof is to show that any eigenstate of $S_z$, in 
other words, any state with a well-defined electron polarization, is 
contained almost completely in either the $+$ or $-$ subspace (note that 
this statement \emph{does not} immediately follow from Theorem 1, in which 
we proved the converse; namely that $+$ or $-$ eigenstates of $H$ are 
almost completely contained within the $\Uparrow$ or $\Downarrow$ 
subspace.)  

\begin{theorem}\label{Theorem::3}
Let $\ket{\Uparrow, \psi}$ be an arbitrary state in which the electron is 
polarized in the $+z$ direction, i.e. $\ket{\psi}$ is some arbitrary state 
of the nuclear spins.  Then,
\begin{equation}
\bra{\Uparrow,\psi}\Pi_-\ket{\Uparrow,\psi} \leq \epsilon_\Uparrow^2,
\end{equation}
where
$\Pi_-$ is the projector onto the $-$ eigenstates of $H$ and
\begin{equation} \label{Equation::Epsilon2Up}
\epsilon_\Uparrow^2 = 
\frac{\left(E_{0,max}^\Uparrow - E_{0,min}^\Uparrow\right)^2 + V_{max}^2}
{\left(E_{max}^- - E_{0,min}^\Uparrow\right)^2}
\end{equation}
Similarly,
\begin{equation}
\bra{\Downarrow,\psi}\Pi_+\ket{\Downarrow,\psi} \leq \epsilon_\Downarrow^2,
\end{equation}
where $\Pi_+$ is the projector onto the $+$ eigenstates of $H$ and
\begin{equation} \label{Equation::Epsilon2Down}
\epsilon_\Downarrow^2 = 
\frac{\left(E_{0,max}^\Downarrow - E_{0,min}^\Downarrow\right)^2 + V_{max}^2}
{\left(E_{min}^+ - E_{0,max}^\Downarrow\right)^2}
\end{equation}
\end{theorem}

To simplify future analysis, we will use
\begin{equation} \label{Equation::Epsilon}
\epsilon = \max{\{\epsilon_\Uparrow,\epsilon_\Downarrow\}}
\end{equation}
in place of $\epsilon_\Uparrow$ and $\epsilon_\Downarrow$.  
Then Theorem 3 demonstrates that an arbitrary state with a $\pm z$ 
polarized electron has at most an amplitude of $\epsilon$ on the $\mp$ 
subspace.   

Finally, let us consider the time evolution of the state 
$\ket{\Uparrow,\psi}$.  We will show that at high fields, the 
$z$-polarization of a state does not vary substantially from its initial 
polarization.  In other words, at fields $B > B_c$, $T_1$ is infinite 
because the $z$-polarization of the electron does not undergo 
substantial decay.

\begin{theorem}\label{Theorem::4}
Let $U$ be the time evolution operator given by,
\begin{equation}
U = \exp{\left(iHt\right)}.
\end{equation}
Given an arbitrary initial state $\ket{\Uparrow,\psi}$ with the electron 
polarized in the $+z$ direction, the following inequality holds for all 
times $t$,
\begin{equation}
\bra{\Uparrow,\psi}U^\dagger(t) S_z U(t)\ket{\Uparrow,\psi} 
\geq \frac{\hbar}{2}\left(1 - 8\epsilon^2\right).
\end{equation}
Similarly, given an arbitrary initial state $\ket{\Downarrow,\psi}$ with the 
electron polarized in the $-z$ direction, 
\begin{equation}
\bra{\Downarrow,\psi}U^\dagger(t) S_z U(t)\ket{\Downarrow,\psi} 
\leq -\frac{\hbar}{2}\left(1 - 8\epsilon^2\right).
\end{equation}
\end{theorem}

Given that the expressions for $\epsilon^2$ and $\xi_2$ are complicated, 
it is useful to observe their general form.  
If we assume that $\hbar \gamma_S B \gg \hbar^2 A_{max}$,  we 
can expand $\epsilon$ and $\xi_2$ to lowest order in $1/B$.  
To the lowest order in $1/B$ we find that $\epsilon^2 \approx 5\xi_2/4$ 
and
\begin{equation} \label{Equation::Xi2Approx}
\xi_2 \approx \frac{\frac{1}{2}\hbar^4 
A_{max}^2}{\left(\hbar 
(\gamma_S - \gamma_I) B\right)^2} \\
\end{equation}
$\epsilon^2$ and $\xi_2$ are essentially the squares 
of the ratio of the Overhauser energy $~ \hbar^2 A_{max}$ to the 
electron 
Zeeman energy $\hbar \gamma_S B$.  If 
this ratio is small, then the $z$ spin of the electron undergoes only 
small fluctuations on the order of $\epsilon^2$.  Thus, for fields greater 
than $B_c$, decay is suppressed and $T_1$ is consequently infinite.

\subsection{Bound on $T_2$}

Let us now turn our attention to the second relaxation time $T_2$.  $T_2$ 
describes the decay of the in-plane magnetization under a spin echo 
experiment, i.e. the decay of the 
expectation value $\abs{\ensavg{S_+(2t)}}$, where a single spin echo 
$\pi$-pulse is 
applied at time $t$.  The bound 
on $T_2$ is based on the variational principle mentioned above.  
It will be shown that the $+$ and $-$ eigenstates obey very similar 
evolution equations.  Thus, the dephasing which occurs between $+$ and $-$ 
eigenstates is limited by the ``magnitude'' of this difference in 
dynamics.  

We first begin with 
some general observations.  Let $\ket{\phi}$ be a wavepacket formed from 
a linear combination of eigenstates of the Hamiltonian $H$.  Then we can 
write,
\begin{eqnarray}
\ensavg{E} &=& \bra{\phi}H\ket{\phi}\\
&=& \sum_k{\abs{\bkm{\psi_k}{\phi}}^2 E_k}\\
&=& \sum_k{p_{k} E_k} \\
\ensavg{E^2} &=& \bra{\phi}H^2\ket{\phi} \\
&=& \sum_k{\abs{\bkm{\psi_k}{\phi}}^2 E_k^2}\\
&=& \sum_k{p_{k} E_k^2}
\end{eqnarray}
where $\ket{\psi_k}$ is an eigenket of $H$ with eigenvalue $E_k$ and where 
we have defined $p_{k} = \abs{\bkm{\psi_k}{\phi}}^2$.  We can view 
$p_{k}$ as a probability distribution over the eigenstates of $H$ 
induced by the state $\ket{\phi}$.  Then we can describe the root 
mean squared width $\Delta E$ of 
the distribution of energies as,
\begin{eqnarray}
\Delta E^2 &=& \ensavg{E^2} - \ensavg{E}^2 \\
&=& \sum_k{p_k \left(E_k - \ensavg{E}\right)^2}.
\end{eqnarray}
Theorems 5 and 6 will apply this idea to bounding the dynamics of 
$\ket{\phi}$.  

\begin{theorem}\label{Theorem::5}
Let $\ket{\phi}$ be a trial state with energy width $\Delta E$, let
$\Delta E_{max} = \alpha \Delta E$ be some energy width cutoff with 
$\alpha > 1$, and let $M$ be the set of eigenstates of $H$ states such 
that $\abs{E_k - \ensavg{E}} > \Delta E_{max}$.  Then,
\begin{equation}
\sum_{k \in M}{p_k} \leq \frac{1}{\alpha^2}
\end{equation}
\end{theorem}

\begin{theorem}\label{Theorem::6}
Given a state $\phi$ with energy width $\Delta E$,
then for times $t$ which satisfy,
\begin{equation}
t < \frac{\hbar\pi}{2\alpha \Delta E},
\end{equation}
with $\alpha > 1$, the free evolution of $\ket{\phi}$ is given by,
\begin{equation}
U(t)\ket{\phi} = \exp{(i\theta)}\sqrt{1-\lambda^2}\ket{\phi} + 
\lambda\ket{r},
\end{equation}
where $\theta$ is some arbitrary phase, $\ket{r}$ is a residual vector 
orthogonal to $\ket{\phi}$ and the parameter $\lambda$ satisfies the 
inequality,
\begin{equation} \label{Equation::LambdaBound}
\lambda^2 \leq c(\Delta E,t)^2.
\end{equation}
The parameter $c(\Delta E,t)$ is a function of the root mean squared 
energy width $\Delta E$ and the time $t$ such that,
\begin{equation}
c(\Delta E, t)^2 = 
1-\left(1-\frac{1}{\alpha^2}\right)\cos{\left(\frac{\alpha\Delta 
E t}{\hbar}\right)}+\frac{1}{\alpha^2},
\end{equation}
and $\alpha$ is a parameter that can be chosen to optimize the resulting bound on $T_2$.
\end{theorem}
Using Theorem 6, it is possible to bound the dynamics of any 
initial state $\ket{\phi}$ by evaluating its energy width, $\Delta E$.  
We will come back to the choice of $\alpha$ later when we 
evaluate the $T_2$ bound for specific systems in Sec. III.

We now turn to the problem of using the results of Theorems 5 and 
6 to place bounds on $T_2$ for our system of an electron spin 
coupled to a lattice of nuclear spins.  We assume that we start with 
a thermally equilibrated ensemble of nuclear spins.  A 
spin echo experiment is 
begins with an initial $\pi/2$ pulse denoted by the operator,
\begin{equation}
R_{\pi/2} = \left(
\ket{\Rightarrow}\bra{\Uparrow}+\ket{\Leftarrow}\bra{\Downarrow}
\right)\tensorm\id_I
\end{equation}
where 
\begin{eqnarray}
\ket{\Rightarrow} &=& 
\frac{1}{\sqrt{2}}\left(\ket{\Uparrow}-\ket{\Downarrow}\right) \\
\ket{\Leftarrow} &=& 
\frac{1}{\sqrt{2}}\left(\ket{\Uparrow}+\ket{\Downarrow}\right).
\end{eqnarray}
The system then 
undergoes free evolution for a time $t$.  This free evolution is followed 
by a $\pi$ pulse, denoted by the operator
\begin{equation} 
R_\pi = \left(\ket{\Downarrow}\bra{\Uparrow} + 
\ket{\Uparrow}\bra{\Downarrow}\right)\tensorm\id_I,
\end{equation}
which flips the electron spin.  Finally, there is a second period of free 
evolution for a time $t$.  This sequence is described by the unitary 
spin echo evolution operator, $\tilde{U}$, 
\begin{equation}
\tilde{U} = U(t) R_\pi U(t).
\end{equation}
The decay function $v(2t)$ is given by the 
magnitude of the in-plane magnetization after this sequence.  

At high fields, it is possible to cool the system well below the Zeeman 
energy of the electron, thereby approximately polarizing the electron in 
the $-z$ direction.  However, because the Zeeman energy of a nucleus is 
$10^3$ smaller than that of the electron, the nuclei will, in general, be 
highly thermalized \cite{deSousa:03B1}.
For the purposes of our analysis, we can then assume that in the 
initial state 
of the system, the electron is polarized in the $-z$ direction (i.e. 
aligned with the field) and is uncoupled from the nuclei, which are in a 
completely mixed state.  The density matrix for such a system is given by
\begin{equation} \label{Equation::Rho0Def}
\rho_0 = k\ket{\Downarrow}\bra{\Downarrow}\tensorm\id_I
\end{equation}
where $k$ is some normalization constant which we will drop for 
convenience.  Thus, we need to calculate the quantity
\begin{equation} 
v(2t) = \abs{\trace{\left[R^\dagger_{\pi/2}\tilde{U}^\dagger S_+ 
\tilde{U} R_{\pi/2}\rho_0\right]}}.
\end{equation}
Using the cyclic property of the trace and \refeqn{Rho0Def}, we can 
simplify this expression to
\begin{eqnarray}
v(2t) 
&=& \abs{\trace{\left[\tilde{U}^\dagger S_+ 
\tilde{U} R_{\pi/2}\rho_0 R^\dagger_{\pi/2} \right]}}\\
&=& \frac{1}{2}\abs{\trace{\left[\tilde{U}^\dagger S_+ 
\tilde{U}(\id_S + R_\pi)\tensorm \id_I\right]}}\\
&=& 
\frac{1}{2}\abs{\trace{\left[\tilde{U}^\dagger S_+ 
\tilde{U}\right]} +
\frac{1}{2}\trace{\left[\tilde{U}^\dagger S_+ 
\tilde{U}R_\pi\right]}} \\
\label{Equation::v2t}
&=& \frac{1}{2}\abs{\trace{\left[\tilde{U}^\dagger S_+ 
\tilde{U}R_\pi\right]}}
\end{eqnarray}
where $\id_S$ is the identity operator on the electron spin.

If we evaluate the trace in \refeqn{v2t} using the eigenbasis of $H$, we 
can separate the contributions of the $+$ and $-$ eigenstates,
\begin{eqnarray}
v(2t)
&=& \frac{1}{2}\abs{\trace{\left[\tilde{U}^\dagger S_+ 
\tilde{U}R_\pi\right]}} \\
&=& \frac{1}{2}\abs{
\sum_i{\bra{\psi_i^-}\tilde{U}^\dagger S_+ 
\tilde{U}R_\pi\ket{\psi_i^-}} +
\sum_i{\bra{\psi_i^+}\tilde{U}^\dagger S_+ 
\tilde{U}R_\pi\ket{\psi_i^+}} }\\
\label{Equation::v2tSep}
&\geq& \frac{1}{2}\abs{
\sum_i{\bra{\psi_i^-}\tilde{U}^\dagger S_+ 
\tilde{U}R_\pi\ket{\psi_i^-}}} - \frac{1}{2}\abs{
\sum_i{\bra{\psi_i^+}\tilde{U}^\dagger S_+ 
\tilde{U}R_\pi\ket{\psi_i^+}} }
\end{eqnarray}
Using Theorem 1 and Theorem 4, we find, after some tedious 
algebra, that the contribution of the $+$ eigenstates is negligible,
\begin{equation} \label{Equation::TracePlus}
\abs{\sum_i{\bra{\psi_i^+}\tilde{U}^\dagger S_+ 
\tilde{U}R_\pi\ket{\psi_i^+}}} \leq 
2\epsilon\sqrt{\xi_2}+4\epsilon^2+\xi_2.
\end{equation}
If we let 
$\ket{\psi^-_i} = \ket{\Uparrow,\psi_{\Uparrow,i}^-} + 
\ket{\Downarrow,\psi_{\Downarrow,i}^-}$ be an eigenstate in the $-$ 
subspace,  
we can also simplify the contribution of the $-$ states,
\begin{equation} \label{Equation::TraceMinus}
\abs{\sum_i{\bra{\psi_i^-}\tilde{U}^\dagger S_+ 
\tilde{U}R_\pi\ket{\psi_i^-}}} \geq 
\abs{\sum_i{\bra{\Uparrow,\psi_{\Downarrow,i}^-}\tilde{U}^\dagger S_+ 
\tilde{U}R_\pi\ket{\Downarrow,\psi_{\Downarrow,i}^-}}} -
2\epsilon\sqrt{\xi_2}-4\epsilon^2\xi_2.
\end{equation}
Combining \refeqto{v2tSep}{TracePlus} we obtain
\begin{equation}
\label{Equation::vt}
v(2t) \geq
\frac{1}{2}\abs{\sum_i{\bra{\Uparrow,\psi_{\Downarrow,i}^-}\tilde{U}^\dagger 
S_+ \tilde{U}R_\pi\ket{\Downarrow,\psi_{\Downarrow,i}^-}}} -
2\epsilon\sqrt{\xi_2}-2\epsilon^2-\frac{1}{2}\xi_2-2\epsilon^2\xi_2
\end{equation}

The problem of evaluating $v(2t)$ has been reduced to determining 
the time evolution of the states $\ket{\Downarrow,\psi_{\Downarrow,i}^-}$ 
and $\ket{\Uparrow,\psi_{\Downarrow,i}^-}$.  The physical 
interpretation of the derivation of \refeqn{vt} is that we
have separated out terms which primarily contribute to dephasing processes 
(i.e., the terms in the sum over i) from terms due to the longitudinal 
relaxation of the electron (i.e., the remaining terms).  These 
longitudinal terms are on the order of $\xi_2$ or $\epsilon^2$ since 
this is the magnitude of longitudinal relaxation demonstrated in 
Theorem 4.

We must next evaluate the sum in \refeqn{vt}.  It is clear that under 
free evolution, 
$\ket{\Downarrow,\psi_{\Downarrow,i}^-}$ 
simply acquires an overall phase (minus some small residual), since it is 
the projection of an eigenstate of $H$ onto the $\Downarrow$ 
subspace.  This statement is formalized in Theorem 7.  
A similar idea can be used to characterize the evolution of 
$\ket{\Uparrow,\psi_{\Downarrow,i}^-}$.
Because $\ket{\Uparrow,\psi_{\Downarrow,i}^-}$ is not necessarily 
close to an eigenstate of $H$, its evolution can be 
complicated and may not in general be characterized simply by an overall 
phase.  If we could calculate the energy width of 
$\ket{\Uparrow,\psi_{\Downarrow,i}^-}$, then we could use Theorem 6 to 
bound its dynamics.  However, because the explicit form of the nuclear 
state $\ket{\psi_{\Downarrow,i}^-}$ is not known, we must use a more 
subtle method of obtaining the energy width.  Theorem 8 shows 
that a trial state $\ket{\phi}$ can be constructed with energy 
width given by a quantity $\norm{W}_2$ that is determined by 
$A_{2,max}, E_{max}^-,$ and $E_{0,max}^\Downarrow$.  It is also shown 
that 
$\ket{\Uparrow,\psi_{\Downarrow,i}^-}$ is the projection of $\ket{\phi}$ 
onto the 
$\Uparrow$ subspace.  Using this fact, Theorem 9 then bounds the 
evolution of 
$\ket{\Uparrow, \psi_{\Downarrow,i}^-}$.  We now present these three 
results.

\begin{theorem}\label{Theorem::7}
Let $\ket{\psi^-_i} = \ket{\Uparrow,\psi_{\Uparrow,i}^-} + 
\ket{\Downarrow,\psi_{\Downarrow_i}^-}$ be an eigenstate of $H$ in the $-$ 
subspace.  Then
\begin{equation}
U\ket{\Downarrow,\psi_{\Downarrow,i}^-} = e^{i\omega_{\Downarrow,i}^- 
t}\ket{\Downarrow,\psi_{\Downarrow,i}^-} + \ket{r_{\Downarrow,i}^-}
\end{equation}
where
\begin{equation}
\ket{r_{\Downarrow,i}^-} = \left(e^{i\omega_{\Downarrow,i}^- 
t}-U\right)\ket{\Uparrow,\psi_{\Uparrow,i}^-}.
\end{equation}
\end{theorem}

Again, because $\ket{\Downarrow,\psi_{\Downarrow,i}^-}$ is close to an 
eigenstate of $H$, its evolution will essentially be characterized by an 
overall phase plus a small residual.  Because 
$\ket{\Uparrow,\psi_{\Downarrow,i}^-}$ is not necessarily close to an 
eigenstate of $H$, we cannot use the same reasoning to characterize its 
evolution.  Instead, we show that 
$\ket{\Uparrow,\psi_{\Downarrow,i}^-}$ is close to a trial 
state $\ket{\phi_i}$ whose evolution we can characterize.

\begin{theorem}\label{Theorem::8}
There exists a trial state $\ket{\phi_i}$ such that
\begin{equation}
\Pi_\Uparrow \ket{\phi_i} = \ket{\Uparrow,\psi_{\Downarrow,i}^-}
\end{equation}
and
\begin{equation}
\bra{\phi_i}H^2\ket{\phi_i} - \bra{\phi_i}H\ket{\phi_i}^2 \leq 
\norm{W}_2^2
\end{equation}
where
\begin{equation}
\norm{W}_2 \leq \frac{\hbar^4}{4\left(E_{max}^- + 
E_{0,max}^\Downarrow\right)}A_{2,max}(1+\epsilon)
\end{equation}
\end{theorem}

According to Theorem 8, the trial state $\ket{\phi_i}$ has an energy width 
of $\norm{W}_2$.  Using Theorem 6, we know that for $t < \hbar \pi / 2 
\norm{W}_2$ the evolution of $\ket{\phi_i}$ will then be well-described by 
phase evolution with some arbitrary phase denoted by 
$\omega_{\Uparrow,i}^- t$ (corresponding to $\theta$ in 
Theorem 4) plus some small residual.  Because $\ket{\phi_i}$ is 
close to the state $\ket{\Uparrow,\psi_{\Downarrow,i}^-}$, Theorem 9 
now uses 
the same reasoning as in Theorem 7 to characterize the evolution of 
$\ket{\Uparrow,\psi_{\Downarrow,i}^-}$.

\begin{theorem}\label{Theorem::9}
Let $\ket{\psi_i^-} = \ket{\Uparrow,\psi_{\Uparrow,i}^-} + 
\ket{\Downarrow,\psi_{\Downarrow,i}^-}$ be an eigenstate of $H$ in the $-$ 
subspace.  Then for $t < \hbar \pi / 2\norm{W}_2$,
\begin{equation}
U\ket{\Uparrow,\psi_{\Downarrow,i}^-} = e^{i\omega_{\Uparrow,i}^- 
t}\sqrt{1-\lambda_i^2}\ket{\Uparrow,\psi_{\Downarrow,i}^-} + 
\lambda_i\ket{r_{\Uparrow,i}^-} + \left(e^{i\omega_{\Uparrow,i}^- 
t}-U\right)\ket{\Downarrow,\phi_{\Downarrow,i}}
\end{equation}
where the following inequalities are satisfied,
\begin{equation}
\label{Equation::InEq1}
\lambda_i^2 \leq c(\norm{W}_2,t)^2,
\end{equation}
\begin{equation}
\label{Equation::InEq2}
\bkm{\Downarrow,\phi_{\Downarrow,i}}{\Downarrow,\phi_{\Downarrow,i}} 
\leq \epsilon^2, 
\end{equation}
\begin{equation}
\label{Equation::InEq3}
\abs{\bkm{r_{\Uparrow,i}^-}{\Uparrow,\psi_{\Downarrow,i}^-}} \leq 
\epsilon.
\end{equation}
\end{theorem}

Using Theorems 7 and 9, we obtain the following results for the 
evolution of $\ket{\Downarrow,\psi_{\Downarrow,i}^-}$ and 
$\ket{\Uparrow,\psi_{\Downarrow,i}^-}$ under the spin echo experiment:
\begin{align}
\label{Equation::SpinEchoDown}
\tilde{U}\ket{\Downarrow,\psi_{\Downarrow,i}^-} \,=&\,
e^{i(\omega_{\Downarrow,i}^-+\omega_{\Uparrow,i}^-) t}
\sqrt{1-\lambda_i^2}\ket{\Uparrow,\psi_{\Downarrow,i}^-} + 
e^{i\omega_{\Downarrow,i}^-t}\lambda_i\ket{r_{\Uparrow,i}^-} \\
&+e^{i\omega_{\Downarrow,i}^-t}\left(e^{\omega_{\Uparrow,i}^-t}-U\right)
\ket{\Downarrow,\phi_{\Downarrow,i}^-}+ U 
R_\pi\ket{r_{\Downarrow,i}^-} 
\notag
\end{align}
\begin{align}
\label{Equation::SpinEchoUp}
\tilde{U}\ket{\Uparrow,\psi_{\Downarrow,i}^-}\, =&\,
e^{i(\omega_{\Downarrow,i}^-+\omega_{\Uparrow,i}^-) t}
\sqrt{1-\lambda_i^2}\ket{\Downarrow,\psi_{\Downarrow,i}^-} + 
\lambda_i U R_\pi\ket{r_{\Uparrow,i}^-} \\
&+\left(e^{i\omega_{\Uparrow,i}^- t}U R_\pi-U R_\pi 
U\right)
\ket{\Downarrow,\phi_{\Downarrow,i}^-} + 
e^{i\omega_{\Uparrow,i}^- t}\sqrt{1-\lambda_i^2} 
\ket{r_{\Downarrow,i}^-}\notag.
\end{align}
One effect of the spin echo evolution is especially important to note.  If 
we compare the free evolution of $\ket{\Downarrow,\psi^-_{\Downarrow,i}}$ 
and $\ket{\Uparrow,\psi^-_{\Downarrow,i}}$, we find that they acquire 
phases of $\omega_{\Downarrow,i}^- t$ and 
$\omega_{\Uparrow,i}^- t$, respectively.  Thus, there is a 
relative phase difference between the two evolutions, which would lead to 
a fast dephasing time $T_2^*$ when we sum over $i$ in \refeqn{vt} to 
account for our initial, incoherent thermal distribution.  However, when 
the spin echo experiment is performed, \refeqs{SpinEchoDown}{SpinEchoUp} 
show that both states evolve with the phase 
factor of $(\omega_{\Downarrow,i}^- + \omega_{\Uparrow,i}^-)t$, 
leading to no relative phase difference.  As expected, the inhomogeneous 
dephasing due to an incoherent distribution over initial states is removed 
by the spin echo experiment.

Finally, we can use \refeqs{SpinEchoDown}{SpinEchoUp} to obtain,
\begin{align} \label{Equation::USplusU}
\bra{\Downarrow,\psi_{\Downarrow,i}^-}
\tilde{U}^\dagger S_+ \tilde{U}\ket{\Uparrow,\psi_{\Downarrow,i}^-} 
\geq&
1-\xi_2-\epsilon\, c(\norm{W}_2,t)-2c(\norm{W}_2,t)^2.
\end{align}

Combining \refeqn{vt} and \refeqn{USplusU}, we obtain as a final result
\begin{equation}
v(2t) \geq 
\frac{1}{2}\left(1-2\xi_2-4\epsilon^2-4\epsilon\sqrt{\xi_2}-4\epsilon^2\xi_2-\epsilon\, 
c(\norm{W}_2,t)-2 c(\norm{W}_2,t)^2\right)
\end{equation}
for $t < \hbar\pi / 2\alpha \norm{W}_2$.  

Though this bound is rigorous, it is somewhat complicated.  In practice, 
numerical evaluation of the expressions for $\epsilon, \xi_2,$ 
and $\norm{W}_2$ can be greatly simplified if we assume that $B \gg B_c$, 
as in \refeqn{Xi2Approx}.  In this case, we can also use the approximation 
that
\begin{equation}
\norm{W}_2 \approx W_{max},
\end{equation}
where
\begin{equation}\label{Equation::Wmax}
W_{max} = \frac{\hbar^4 A_{2,max}}{4(\hbar \gamma_S B)}.
\end{equation}
It should be emphasized that this approximation ($B \gg B_c$) is not 
intrinsic to our proof and is used only to make our results more compact 
and easier to present.  If a strict bound were desired, we could 
drop this approximation and keep only the weaker condition that $B > B_c$.

To obtain a bound on $T_2$, we set $v(2t) = 1/2e$,
\begin{equation}
c(\norm{W}_2,t)^2 + \epsilon\, c(\norm{W}_2,t) + 
\xi_2+2\epsilon^2+2\epsilon\sqrt{\xi_2}+2\epsilon^2\xi_2+\frac{1}{2e}
-\frac{1}{2}=0
\end{equation}
Let $c_{max}$ be the positive root of this equation.  Then we solve for 
$t_b$ in terms of $\alpha$ such that $c(\norm{W}_2,t_b) = c_{max}$,
\begin{equation} \label{Equation::TBound}
t_b = \max_{\alpha}{\left\{\frac{\hbar}{\Delta E 
\alpha}\cos^{-1}{\left(\frac{1-c_{max}^2+\frac{1}{\alpha^2}}
{1-\frac{1}{\alpha^2}}\right)}\right\}}.
\end{equation}
This value of $t_b$ will be a lower bound on $T_2$.

\section{Application to Physical Systems}
\label{Section::Application}
To demonstrate how our bounding method can be used in practice, we apply 
it here to three systems of relevance to quantum information 
processing\cite{Loss:98A,Vrijen:99A}: (1) a donor impurity in Si, (2) a 
donor impurity in GaAs, and (3) a GaAs quantum dot.
Table 1 shows the important physical parameters for these systems.  To obtain a lower bound 
on $T_2$ for each system, we must evaluate \refeqn{TBound} for the given system's parameters.  
Because we are free to choose the parameter $\alpha$ in Theorem 6 with only the restriction that $\alpha > 1$, 
we can select $\alpha$ to maximize $t_b$.  In each of the following examples, this optimization 
procedure was carried out to obtain the largest possible $t_b$.

Consider first the case of a phosphorous impurity atom in a lattice
of natural silicon (which contains $4.67 \%$ of the spin-$1/2$ 
$^{29}$Si isotope).  Spin coherence is affected by hyperfine 
coupling between the electron and
nuclei and dipolar coupling between nuclei.  The hyperfine coupling 
constants
are given by the equation \cite{Paget:77B,deSousa:03B1},
\begin{equation}
A_j = \frac{8\pi}{3}\gamma_S \gamma_I \hbar \abs{\Psi(\v{x}_j)}^2,
\end{equation}
where $\v{x}_j$ is the position of nuclear spin $j$ and $\Psi$ is a
hydrogenic orbital for the impurity electron.  Using our bounding 
technique,
we can determine when the $S_+ I_- + S_- I_+$ terms in 
the Hamiltonian can be safely neglected as the primary source for 
electron spin decoherence.  
The coherence time due to dipole-dipole coupling of nuclei for 
a sample of natural Si in the presence of an external field aligned 
with the $\left[1 1 1\right]$ direction has been estimated to be $T_2 
\sim 0.65 
\,\mathrm{msec}$ \cite{deSousa:03B1}.  This value is essentially independent
of the magnetic field strength.  In contrast, Figure 2 shows that our 
lower bound is directly proportional to $B$.  
Using the data from Figure 2, we find that at 
approximately $B=3.0\,\mathrm{Tesla}$, decoherence due to the dipolar 
nuclear coupling 
dominates that due to the hyperfine interaction.  Thus, for this 
system, at fields above $3.0\, \mathrm{Tesla}$ it is safe to 
neglect the 
$S_+ I_- + S_- I_+$ terms in the Hamiltonian as sources for 
decoherence.

Next, we investigate the case of a donor impurity in GaAs and a 
GaAs quantum dot ($20$ nm radius, $10$ nm thickness).  
Ga and As nuclei are both spin $3/2$.  We can easily extend our spin-$1/2$ formalism to 
include higher spin nuclei if we note that a spin-$3/2$ particle can be 
treated as a composite system of three spin-$1/2$ particles.  However, we 
also note that the Ga and As nuclei will have different gyromagnetic 
ratios
$\gamma_{^{71}Ga} =  8.16\times10^7\, (\mathrm{s\,Tesla})$,
$\gamma_{^{69}Ga} =  6.42\times10^7\, (\mathrm{s\,Tesla})$, and
$\gamma_{^{75}As} =  4.58\times10^7\, (\mathrm{s\,Tesla})$.  As a 
result
the Zeeman energy of the electron can be exchanged with the nuclear 
Zeeman energy, providing a new channel for $T_1$ relaxation.  
In order to apply our bound to the GaAs systems, we therefore must make 
the 
assumption that at high fields nuclear polarization is not transferred 
between nuclei of different types.  We then let $\gamma_I = 
\gamma_{^{71}Ga}$, and apply our proof as before.  
Again, the important parameters are shown in Table 1.  Figure 3 shows 
the lower bound on $T_2$ as a function of $B$ for the donor impurity 
in a GaAs lattice (solid line) and a GaAs quantum dot (dot-dashed line).  
A lower bound for $T_2$ is provided for fields greater than $\sim 7.5\, 
\mathrm{Tesla}$.  For both systems, our method also shows that $T_1$ is 
infinite above a critical field of $\sim 2.25\, \mathrm{Tesla}$.  
Previous studies have shown that nuclear spectral 
diffusion causes decoherence on a timescale of $\sim 10 
\,\mathrm{\mu sec}$ for both these GaAs systems \cite{deSousa:03B1,deSousa:04X}.  
Hence, Figure 3 shows that for magnetic fields greater than approximately 
$12.5 \, 
\mathrm{Tesla}$, 
nuclear spectral diffusion will be the dominant 
decoherence mechanism 
and the $S_+ I_- + S_- I_+$ terms in the Hamiltonian can be 
neglected.

\section{Discussion and Conclusions}
\label{Section::Conclusions}
It is useful to assign a physical significance to the quantities that 
appear in our bounds.  For instance, the relevant quantity in the $T_1$ 
bound is given alternatively by $\epsilon^2$ or $\xi_2$, both of which
quantify the 
mixing of the spin-up and spin-down eigenstates of $H_0$.  These 
quantities scale as the square of the ratio of the Overhauser energy 
$\hbar^2 A_{max}$ 
and the electron Zeeman splitting, $\hbar\gamma_S B$.  In the language of 
perturbation theory, the large energy denominator suppresses mixing of 
spin-up and spin-down states via the perturbation $V$.

On the other hand, the operator which sets the scale for $T_2$ is $W$, which 
can be written as,
\begin{equation} \label{Equation::WDef2}
W = \Pi_\Uparrow\left(R_\pi V\frac{1}{E_i^--H_0}V R_\pi - 
V\frac{1}{E_i^- + H_0}V\right)\Pi_\Uparrow.
\end{equation}
This expression motivates us to identify the operator $W$ with difference 
in effective Hamiltonians felt by electron spin-up and electron spin-down 
states.  For instance, the second term in this equation describes the 
Hamiltonian experienced by an 
electron spin-up state scattered via two applications of $V$ into a second 
electron spin-up state.  On the other hand, the first term applies the 
same scattering process to an electron spin-down state.  Taking the 
difference of these two terms allows cancellation to occur, reducing the 
magnitude of $W$, and corresponding to the refocusing of spin states 
accomplished by the spin echo experiment.  If no spin echo were 
applied, then the first of these terms would be absent, and the magnitude 
of $W$ would be substantially larger, leading to a faster decay of 
coherence.

In summary, we have used a completely non-perturbative approach to 
place a rigorous lower bounds on the $T_1$ and $T_2$ 
coherence times of an electron spin coupled via the contact hyperfine 
interaction to a lattice of nuclear spins.  For $T_1$, we have shown 
that for $B > B_c$, the $z$-polarization of the electron is nearly 
conserved by time evolution; hence $T_1$ is infinite.  For $T_2$, we 
have obtained an analytic expression for a lower bound $t_b$, having 
explicitly taken into account the effects of a spin echo experiment, which 
would be required to remove any additional inhomogeneous broadening.
This analysis holds above some critical field satisfying the 
high field condition, $B 
> B_c$.  We have also assumed that at the start of 
an experiment, the electron is polarized and uncoupled from the 
completely mixed nuclear initial state; similar results can be derived  
for any highly mixed nuclear initial state.
In general, this method can be 
used to determine when the contact hyperfine interaction can be safely 
neglected as a direct mechanism for electron spin decoherence.  We have 
demonstrated the utility of the bounds for a donor impurity in Si, and 
-within the approximation that spontaneous nuclear polarization transfer 
does not occur- for a donor impurity in GaAs, and a GaAs quantum dot.

Future directions for this investigation are two-fold.  First, 
numerical simulations indicate that the 
coherence time $T_2$ due to hyperfine coupling only might, in fact, be 
substantially higher than the 
lower bound given here, possibly even infinite \cite{Shenvi:04X2}.  The 
primary contribution 
to $T_2$ comes from the diagonal matrix elements of $W$, which 
corresponds to the self-energy of the perturbation.  Removing this 
term would greatly 
improve the lower bound on $T_2$; unfortunately, it is unclear how this 
term can be suppressed.  Also, we have not taken advantage of the 
substantial symmetries present in the system \cite{Gaudin:76,Shenvi:04X2}.  
Taking advantage of these symmetries in our analysis could 
dramatically increase our lower bound on $T_2$.
Second, this method might be 
extended to yield bounds on the fidelity of single qubit operations on 
the electron spin, which would be important for any solid state 
quantum computing applications.  
Finally, it is likely that our method could be extended to any quasi-two 
level system with a significant energy gap that is coupled to a bath 
via an off-diagonal perturbation.  It would be useful to generalize the 
method 
so that it could be applied to other systems of interest to quantum 
computation.

\begin{table}
\begin{tabular}{|l|c|c|c|}
\hline
System & $B_c (\mathrm{Tesla})$ & 
$\hbar / W_{max} (\mathrm{s / Tesla})$ \\ \hline
Si:donor & $9.2\times10^{-4}$ & $1.31\times 
10^{-3}$ \\ \hline
GaAs:donor & $2.24$ & $1.45\times 
10^{-5}$ \\ \hline
GaAs dot & $2.26$ & $9.67\times 
10^{-6}$ \\ \hline
\end{tabular}
\caption{This table lists the relevant parameters for our 
calculations of the lower bounds on $T_1$ and $T_2$ for the electron 
spin.  $B_c$ is given by \refeqn{BCondition} and $W_{max}$ by \refeqn{Wmax}.
The parameters for the Si:donor and GaAs:donor systems 
were calculated using simple hydrogenic orbitals as described in 
\cite{Paget:77B,deSousa:03B1}.  The GaAs quantum dot considered 
had a radius of $20\,\mathrm{nm}$ and a thickness of 
$10\,\mathrm{nm}$.  The parameters for the quantum dot were 
calculated as described in \cite{deSousa:03B1}.  Here $\hbar B / 
W_{max}$ provides an approximate bound for $T_2$ for any field strength $B$.
}
\end{table}

\pagebreak[4]
\begin{figure}
\includegraphics[width=7in,height=6in]{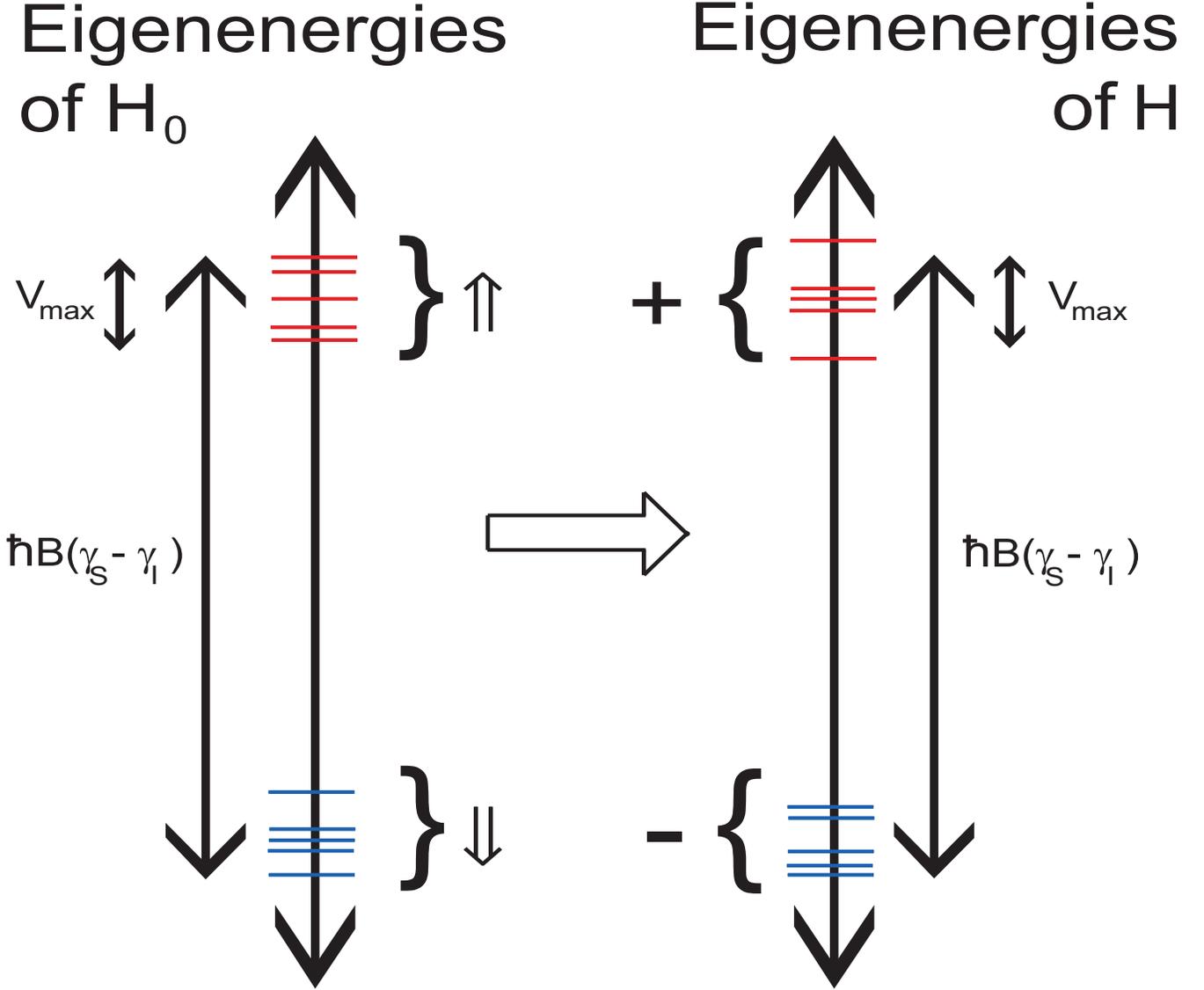}
\caption{
A schematic representation of the eigenspectra of $H_0$ and $H$.  The 
eigenstates of the unperturbed Hamiltonian can be grouped into two 
subspaces, $\Uparrow$ and $\Downarrow$ based on the $z$-polarization of 
the electron.  In Theorems 1 and 2, we demonstrate that the eigenstates of 
the full Hamiltonian $H$ can be grouped into two subspaces, $+$ and $-$, 
based on the approximate $z$-polarization of the electron.  Theorem 3 
further states that the $+$ and $-$ eigenstates fall within the same range 
as the $\Uparrow$ and $\Downarrow$ eigenstates, respectively.
 }
\end{figure}

\pagebreak[4]
\begin{figure}
\includegraphics*{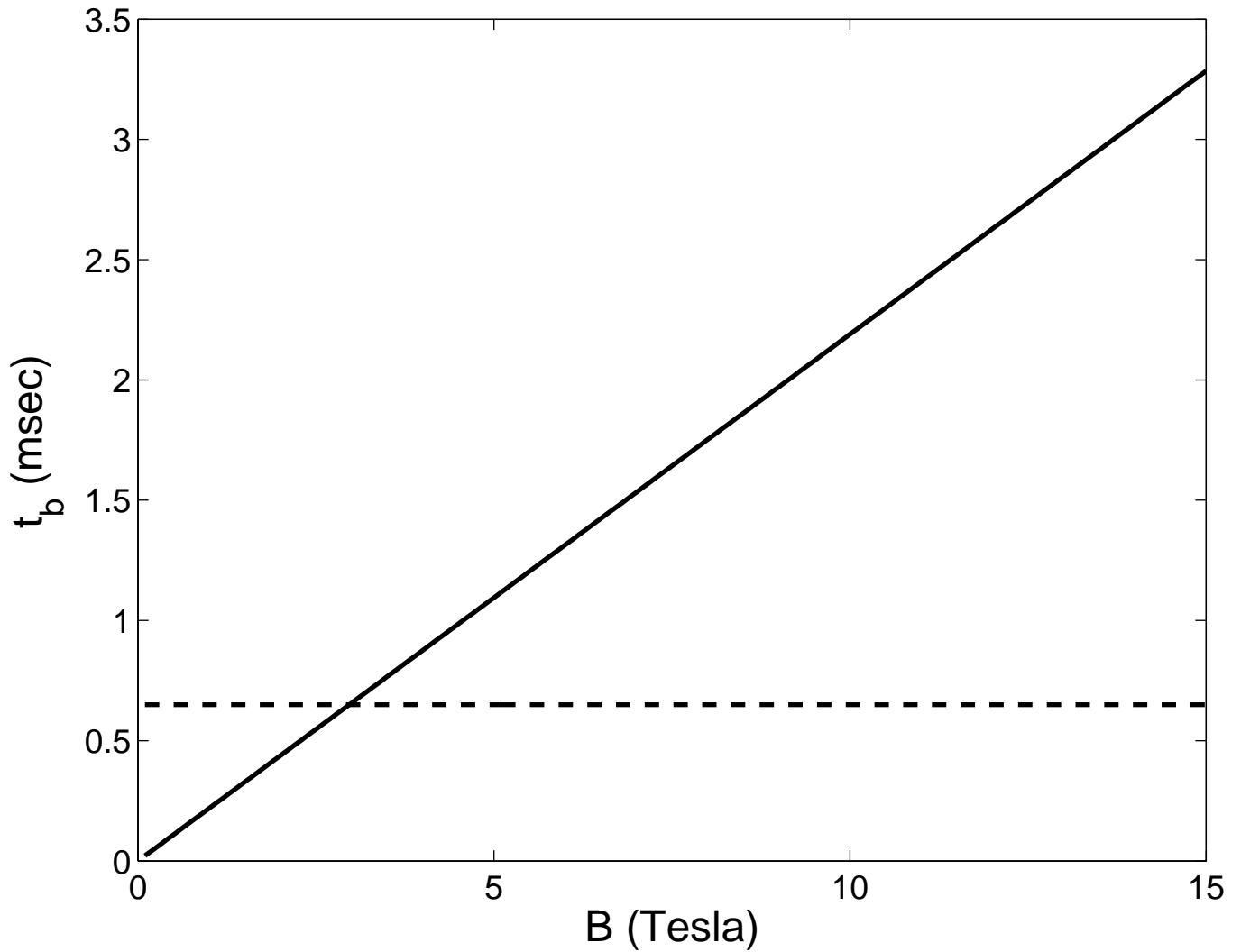}
\caption{
Lower bound on $T_2$ versus magnetic field for a donor impurity in a 
sample of natural Si.  
Note: this plot shows the dependence of the \emph{lower 
bound} on the magnetic field, not the dependence of $T_2$ itself.  
At high magnetic fields $B \gg B_c$ the lower bound scales approximately 
as $B$.  The dashed line shows the predicted timescale for decoherence due 
to nuclear spectral diffusion arising from internuclear dipolar coupling 
\cite{deSousa:03B1}.  Thus, at fields greater 
than $3.0\, \mathrm{Tesla}$, this nuclear spectral diffusion will 
dominate hyperfine 
coupling as a source for decoherence.
}
\end{figure}

\pagebreak[4]
\begin{figure}
\includegraphics{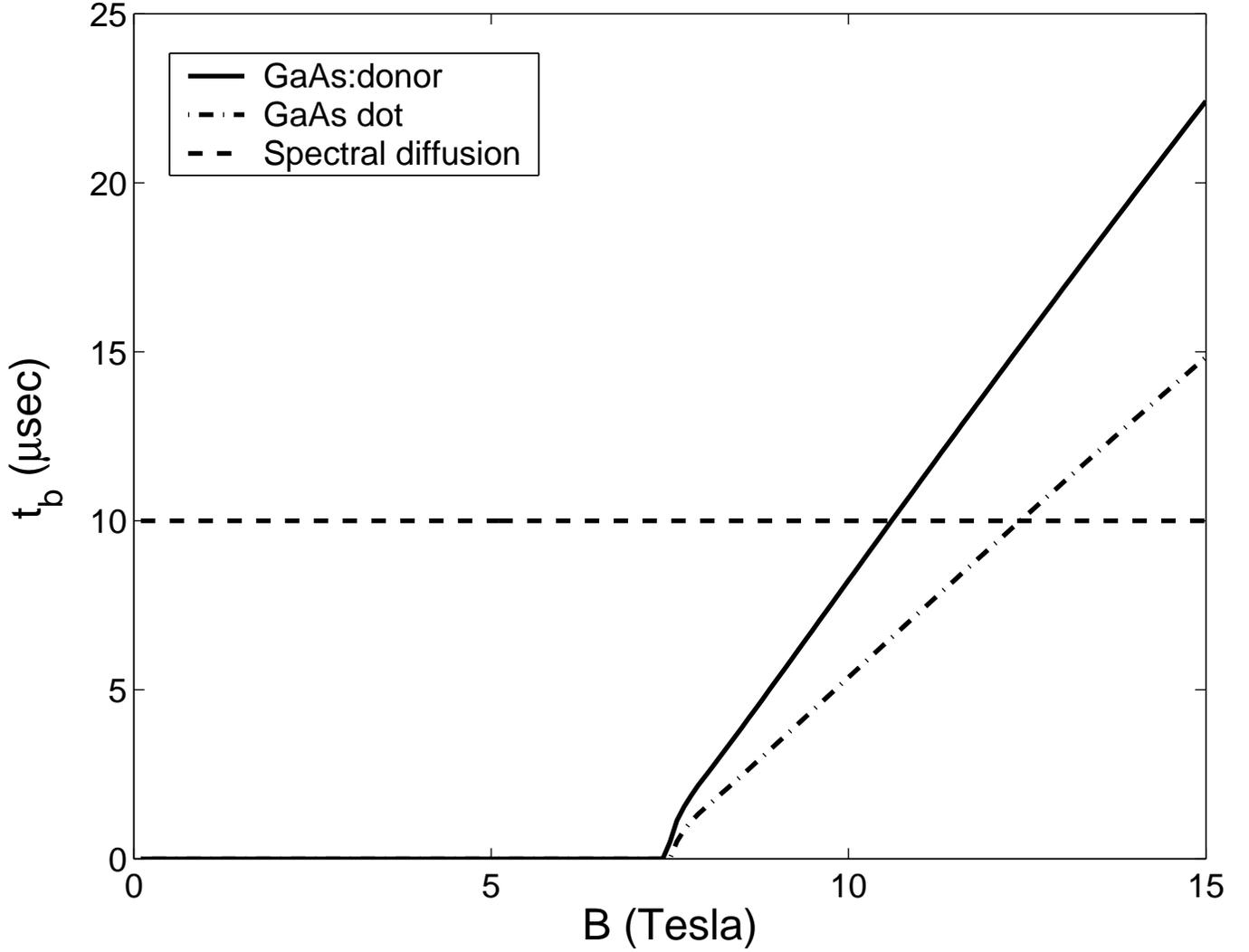}
\caption{
Lower bound on $T_2$ versus magnetic field $B$ for a donor 
impurity in GaAs and a GaAs quantum dot with radius $l_0 = 
20\,\mathrm{nm}$ and thickness $z_0 = 10\,\mathrm{nm}$.  The dashed 
line shows the approximate decoherence time of $\sim 10\, 
\mathrm{\mu sec}$ for both systems due to dipolar-induced nuclear spectral 
diffusion 
\cite{deSousa:03B1,deSousa:04X}.  At fields above $\sim 12.5\, 
\mathrm{Telsa}$, spectral diffusion will dominate hyperfine 
coupling as a source for decoherence.
}
\end{figure}

%\bibliography{Paper9e}
%\bibliographystyle{apsrev}

\pagebreak[4]
\setcounter{page}{1}
\setlength{\voffset}{2.5cm}

\section{Appendix} \label{Section::Proof}
\begin{proof}[Proof of Theorem~\ref{Theorem::1}]
Let $\ket{\psi}$ be an eigenstate of $H$ such that,
\begin{equation}\label{Equation::EigHL}
H\ket{\psi} = E\ket{\psi}.
\end{equation}
Without loss of generality, we can write $\ket{\psi}$ as,
\begin{equation}
\ket{\psi} = \ket{\Uparrow, \psi_\Uparrow} + \ket{\Downarrow, 
\psi_\Downarrow}.
\end{equation}
where
\begin{eqnarray} \label{Equation::PiUparrowPsi}
\ket{\Uparrow, \psi_\Uparrow} &=& \Pi_\Uparrow \ket{\psi},\\
\ket{\Downarrow, \psi_\Downarrow} &=& \Pi_\Downarrow \ket{\psi},
\end{eqnarray}
Recall that $\Pi_\Uparrow, \Pi_\Downarrow$ are the electron spin 
projection operators defined in \refeqs{PiUp}{PiDown}.
Multiplying \refeqn{EigHL} on the left by either $\bra{\Uparrow, 
\psi_\Uparrow}$ or $\bra{\Downarrow, \psi_\Downarrow}$, we 
obtain the two equations,
\begin{eqnarray}
\bra{\Uparrow, \psi_\Uparrow} H_0 \ket{\Uparrow, \psi_\Uparrow} + 
\bra{\Uparrow, \psi_\Uparrow}V 
\ket{\Downarrow, \psi_\Downarrow} 
&=& E \bkm{\Uparrow,\psi_\Uparrow}{\Uparrow,\psi_\Uparrow} \\
\bra{\Downarrow, \psi_\Downarrow} H_0 \ket{\Downarrow, \psi_\Downarrow} + 
\bra{\Downarrow, \psi_\Downarrow}V 
\ket{\Uparrow, \psi_\Uparrow} &=& E 
\bkm{\Downarrow,\psi_\Downarrow}{\Downarrow,\psi_\Downarrow}.
\end{eqnarray}

It is straightforward to set a bound on the quantities
$\bra{\Uparrow, \psi_\Uparrow} H_0 \ket{\Uparrow, \psi_\Uparrow}$ and 
$\bra{\Downarrow, \psi_\Downarrow} H_0 \ket{\Downarrow, \psi_\Downarrow}$ 
since the eigenspectrum of $H_0$ is known (see \refeqs{E0Plus}{E0Minus}).  
Specifically, it can be shown that,
\begin{equation} 
\bra{\Uparrow,\psi_\Uparrow}H_0\ket{\Uparrow,\psi_\Uparrow} \geq 
E_{0,min}^\Uparrow c_\Uparrow^2
\end{equation}
and
\begin{equation}
\bra{\Downarrow,\psi_\Downarrow}H_0\ket{\Downarrow,\psi_\Downarrow} \leq 
E_{0,max}^\Downarrow c_\Downarrow^2,
\end{equation}
where
\begin{eqnarray} 
\label{Equation::CUparrowDef}
c_\Uparrow &=& \sqrt{\bkm{\Uparrow,\psi_\Uparrow}{\Uparrow,\psi_\Uparrow}}\\
\label{Equation::CDownarrowDef}
c_\Downarrow &=& \sqrt{\bkm{\Downarrow,\psi_\Downarrow}{\Downarrow,\psi_\Downarrow}}.
\end{eqnarray}

Here, we pause to introduce several relations involving matrix norms 
which we will make use of in what follows.  Let $X$ be an 
arbitrary Hermitian matrix.  Then following the standard 
definitions, we let $\norm{X}_1$ represent the maximum absolute column 
sum norm
\begin{equation}
\norm{X}_1 = \max_j{\sum_i{\abs{X_{ij}}}}.
\end{equation}
The maximum absolute row sum norm is given by
\begin{equation}
\norm{X}_\infty = \max_i{\sum_j{\abs{X_{ij}}}}.
\end{equation}
Finally, the spectral norm, can be defined as 
\begin{equation} \label{Equation::L2NormDef}
\norm{X}_2 = \max_{\ket{\phi}}{\abs{\bra{\phi}X\ket{\phi}}}.
\end{equation}
When applied to a vector $\norm{\ket{\psi}}_2$ is simply the $L_2$ norm of 
$\ket{\psi}$.  
We will also use two useful relations regarding these 
norms\cite{Higham:92}
\begin{eqnarray}
\label{Equation::Norm2Psi}
\norm{X \ket{\psi}}_2 &\leq& \norm{X}_2 \norm{\ket{\psi}}_2 \\
\label{Equation::Norm2X}
\norm{X}_2 &\leq& \norm{X}_1 \norm{X}_\infty.
\end{eqnarray}

We can use these relations applied to the matrix $V$ to bound the 
remaining quantities in \refeqs{CUparrowDef}{CDownarrowDef}.  
Using \refeqn{Norm2Psi} we obtain
\begin{equation}
\abs{\bra{\Uparrow,\Psi_\Uparrow}V\ket{\Downarrow,\Psi_\Downarrow}} 
\leq c_\Uparrow c_\Downarrow \norm{\Pi_\Uparrow V \Pi_\Downarrow}_{2}.
\end{equation}
Because the absolute column sum and absolute row sum norms of $V$ are 
easy to calculate, we can use them to bound the spectral norm of $V$ 
within a given $L$ subspace using \refeqn{Norm2X}
\begin{eqnarray}
\norm{\Pi_\Uparrow V \Pi_\Downarrow}_{2} &\leq& 
\frac{1}{2}\hbar^2 
\sqrt{A[L] A[N-L+1]} \\
&\equiv& V_{max}
\end{eqnarray}
Similarly,
\begin{equation} \label{Equation::NormV}
\norm{\Pi_\Downarrow V \Pi_\Uparrow}_{2} \leq V_{max}.
\end{equation}
Then we obtain the two bounding inequalities,
\begin{eqnarray} 
\label{Equation::CplusE}
E_{0,min}^{\Uparrow}c_\Uparrow^2 - c_\Uparrow c_\Downarrow V_{max} &\leq& 
E c_\Uparrow^2 \\
\label{Equation::CminusE}
E_{0,max}^{\Downarrow}c_\Downarrow^2 + c_\Uparrow c_\Downarrow V_{max} 
&\geq& E c_\Downarrow^2 \\
\end{eqnarray}
Combining \refeqs{CplusE}{CminusE}, we obtain the inequality
\begin{equation}\label{Equation::xyInequality}
c_\Uparrow^2 + c_\Downarrow^2 + c_\Uparrow c_\Downarrow 
\frac{E_{0,min}^\Uparrow 
- E_{0,max}^\Downarrow}{V_{max}}\geq 0.
\end{equation}
Since the normalization of $\ket{\psi}$ implies that $c_\Uparrow^2 + 
c_\Downarrow^2 = 1$,
some brief algebra then yields
\begin{equation} \label{CUparrowIneq}
c_\Uparrow^4 - c_\Uparrow^2 + \frac{V_{max}^2}{(E_{0,min}^\Uparrow - 
E_{0,max}^\Downarrow)^2} \geq 0.
\end{equation}
Solving the equality for $c_\Uparrow^2$, we obtain the two solutions 
$c_\Uparrow = \xi_1,\xi_2$ where $\xi_1$ and $\xi_2$ are given in 
\refeqn{Xi12Def}.
Thus, using \refeqs{PiUparrowPsi}{CUparrowDef}, we 
have shown that every eigenstate $\ket{\psi_k}$ of $H$ satisfies one of 
the two inequalities,
\begin{equation} \label{Equation::PsiPlus}
\bra{\psi_k}\Pi_\Uparrow\ket{\psi_k} \geq \xi_{1}
\end{equation}
or
\begin{equation} \label{Equation::PsiMinus}
\bra{\psi_k}\Pi_\Uparrow\ket{\psi_k} \leq \xi_{2}
\end{equation}
QED
\end{proof}

\begin{proof}[Proof of Theorem~\ref{Theorem::2}]
We will show only one of these bounds as the rest can be demonstrated 
by identical procedures.  For a $+$ eigenstate, we know from the proof 
of Theorem 1 that 
\begin{eqnarray}
c_\Downarrow &\leq& \sqrt{\xi_2} \\
c_\Uparrow &\geq& \sqrt{\xi_1}.
\end{eqnarray}
Then \refeqn{CplusE} gives us,
\begin{eqnarray} \label{Equation::EkPlus}
E_k^+ &\geq& E_{0,min}^\Uparrow - \sqrt{\frac{\xi_2}{\xi_1}} V_{max}\\
&\geq& E_{0,min}^\Uparrow - V_{max}\\
&=& E_{min}^+.
\end{eqnarray}
Where we have recalled that our high field condition ensures that $\xi_1 > 
\xi_2$ (see \refeqn{Xi12Def}).  The other bounds are derived in exactly 
the same way.  
QED  
\end{proof}

\begin{proof}[Proof of Theorem~\ref{Theorem::3}]
We will demonstrate this theorem for the spin-up case; 
the spin-down case follows the same argument.  Let $\ket{\Uparrow,\psi}$ 
be an arbitrary state with the electron polarized in the $+z$ 
direction.  Since $\ket{\Uparrow,\psi}$ has no spin-down component, its 
expectation value over $V$ is $0$.  Given the eigenspectrum of $H_0$, it 
is straightforward to show that,
\begin{equation}
E_{0,max}^\Uparrow \geq \bra{\Uparrow,\psi}H\ket{\Uparrow,\psi} \geq 
E_{0,min}^\Uparrow.
\end{equation}
Given the matrix norm of the perturbation $V$ in \refeqn{NormV}, it is 
equally straightforward to show that,
\begin{equation} \label{Equation::VarHL}
\bra{\Uparrow,\psi}\left(H - \ensavg{H}\right)^2\ket{\Uparrow,\psi} \leq 
\frac{1}{4}\left(E_{0,max}^\Uparrow - E_{0,min}^\Uparrow\right)^2 + 
V_{max}^2.
\end{equation}
However, we can also write the left hand side of \refeqn{VarHL} as,
\begin{eqnarray}
\bra{\Uparrow,\psi}\left(H-\ensavg{H}\right)^2\ket{\Uparrow,\psi}
&=&
\sum_k{\left(
\abs{\bkm{\Uparrow,\psi}{\psi_k^+}}^2\left(E_k^+ - \ensavg{H}\right)^2 + 
\abs{\bkm{\Uparrow,\psi}{\psi_k^-}}^2\left(E_k^- - \ensavg{H}\right)^2\right)}\\
&=&
\sum_k{\left(
p_k^+\left(E_k^+ - \ensavg{H}\right)^2 +
p_k^-\left(E_k^- - \ensavg{H}\right)^2\right)},
\end{eqnarray}
where 
\begin{equation}
p_k^\pm = \abs{\bkm{\Uparrow,\psi}{\psi_k^\pm}}^2.
\end{equation}
Then using \refeqs{EPlusBound}{EMinusBound}, we can bound this quantity 
as,
\begin{eqnarray} 
\sum_k{
p_k^+\left(E_k^+ - \ensavg{H}\right)^2 +
p_k^-\left(E_k^- - \ensavg{H}\right)^2}
&\geq&
\left(0\right)^2 \sum_k{p_k^+}
+\left(E_{max}^- - E_{0,min}^\Uparrow\right)^2\sum_k{p_k^-}\\
\label{Equation::PkPlusPkMinus}
&=&
\left(E_{max}^- - E_{0,min}^\Uparrow\right)^2
\bra{\Uparrow,\psi}\Pi_-\ket{\Uparrow,\psi}.
\end{eqnarray}
Combining \refeqs{VarHL}{PkPlusPkMinus}, we obtain,
\begin{equation}
\label{Equation::UpPminusUp}
\bra{\Uparrow,\psi}\Pi_-\ket{\Uparrow,\psi}
\leq
\epsilon_\Uparrow^2
\end{equation}
where
\begin{equation}
\epsilon_\Uparrow^2 = 
\frac{\frac{1}{4}\left(E_{0,max}^\Uparrow - E_{0,min}^\Uparrow\right)^2 + 
V_{max}^2}
{\left(E_{max}^- - E_{0,min}^\Uparrow\right)^2}.
\end{equation}
QED
\end{proof}

\begin{proof}[Proof of Theorem~\ref{Theorem::4}]
We will demonstrate the inequality for the initial state 
$\ket{\Uparrow,\psi}$; the spin-down inequality can be proved in a similar 
manner.  Given the initial state $\ket{\Uparrow,\psi}$, the state at time 
$t$ will be given by, 
\begin{equation}
\ket{\psi(t)} = U\ket{\Uparrow,\psi}.
\end{equation}
The amplitude in the electron spin-down subspace after 
time t is given by,
\begin{eqnarray}
\norm{\Pi_\Downarrow U\ket{\Uparrow,\psi}}_2
&=& \max_{\ket{\phi}}\abs{{\bra{\Downarrow,\phi}U\ket{\Uparrow,\psi}}}
\\
&=& \max_{\ket{\phi}}{\abs{\bra{\Downarrow,\phi}\left(\Pi_+ 
+ \Pi_-\right)U\left(\Pi_+ + \Pi_-\right)\ket{\Uparrow,\psi}}} \\
&\leq& 
\max_{\ket{\phi}}{\abs{\bra{\Downarrow,\phi}\Pi_+U \Pi_+\ket{\Uparrow,\psi}}}+
\max_{\ket{\phi}}{\abs{\bra{\Downarrow,\phi}\Pi_-U \Pi_-\ket{\Uparrow,\psi}}}\\
&\leq& 
\max_{\ket{\phi}}{\norm{\Pi_+\ket{\Downarrow,\phi}}_2\cdot\norm{\Pi_+\ket{\Uparrow,\psi}}_2}+
\max_{\ket{\phi}}{\norm{\Pi_-\ket{\Downarrow,\phi}}_2\cdot\norm{\Pi_-\ket{\Uparrow,\psi}}_2}\\
\label{Equation::PiMinusUUp}
&\leq& 2\epsilon
\end{eqnarray}
where the maximization is taken over all normalized nuclear states 
$\ket{\phi}$.  We have also used \refeqn{UpPminusUp} and the fact that $U$ 
and $\Pi_\pm$ commute.  Similarly, we can obtain,
\begin{equation}\label{Equation::PiPlusUUp}
\norm{\Pi_\Uparrow U\ket{\Uparrow,\psi}}_2 \geq \sqrt{1 - 
4\epsilon^2}.
\end{equation}
Using \refeqs{PiMinusUUp}{PiPlusUUp}, we can now bound the z-polarization 
of the electron starting from an initial state which is polarized.  
First, we note that the operator $S_z$ can be written in terms of the 
projection operators $\Pi_\Uparrow$ and $\Pi_\Downarrow$,
\begin{equation}
S_z = \frac{\hbar}{2}\left(\Pi_\Uparrow - 
\Pi_\Downarrow\right)\tensorm\id_I
\end{equation}
Using \refeqn{PiPlusUUp}, we find that if the initial state of the system 
is given by $\ket{\Uparrow,\psi}$, 
then the z-polarization at an arbitrary time $t$ is bounded by,
\begin{eqnarray}
\bra{\Uparrow,\psi}U^\dagger S_z U\ket{\Uparrow,\psi} 
&=&
\frac{\hbar}{2}\left(
\bra{\Uparrow,\psi}U^\dagger \Pi_\Uparrow U\ket{\Uparrow,\psi} -
\bra{\Uparrow,\psi}U^\dagger \Pi_\Downarrow U\ket{\Uparrow,\psi}\right)\\
&\geq& \frac{\hbar}{2}\left(1 - 8\epsilon^2\right).
\end{eqnarray}
QED
\end{proof}

\begin{proof}[Proof of Theorem~\ref{Theorem::5}]
From the statement of Theorem 5, we have defined $\Delta E_{max} = \alpha 
\Delta E, \alpha > 1$.  We have also defined the set of eigenstates $M$ 
such that for every eigenstate $k$ in $M$, we have $\abs{E_k - \ensavg{E}} 
> \Delta E_{max}$.  Then we can write
\begin{eqnarray} \label{Equation::pMBound}
\ensavg{\left(\Delta E\right)^2} &=&  
\sum_k{p_k \left(E_k - \ensavg{E}\right)^2}\\
&=& \sum_{k\in \emph{M}}{p_k \left(E_k - \ensavg{E}\right)^2}+
\sum_{k\notin \emph{M}}{p_k \left(E_k - \ensavg{E}\right)^2}\\
&\geq&  
\sum_{k\in \emph{M}}{p_k \left(E_k - \ensavg{E}\right)^2}\\
&\geq&  
\sum_{k\in \emph{M}}{p_k \left(\Delta E_{max}\right)^2}\\
&=&  
\left(\Delta E_{max}\right)^2\sum_{k\in \emph{M}}{p_k}\\
\end{eqnarray}
So,
\begin{equation}
\sum_{k\in \emph{M}}{p_k} \leq \frac{1}{\alpha^2}.
\end{equation}
QED
\end{proof}

\begin{proof}[Proof of Theorem~\ref{Theorem::6}]
\begin{eqnarray}
\abs{\bra{\phi}U\ket{\phi}} 
&=& \abs{\sum_k{p_k \exp{\frac{i E_k t}{\hbar}}}} \\
&=& \abs{\sum_{k\notin \emph{M}}{p_k \exp{\frac{i E_k t}{\hbar}}} 
+ \sum_{k\in \emph{M}}{p_k \exp{\frac{i E_k t}{\hbar}}}} \\
&\geq& \abs{\sum_{k\notin \emph{M}}{p_k \exp{\frac{i E_k t}{\hbar}}}} 
- \abs{\sum_{k\in \emph{M}}{p_k \exp{\frac{i E_k t}{\hbar}}}} \\
\label{Equation::OverallE}
&\geq& \abs{\exp{\frac{i \ensavg{E} t}{\hbar}}\sum_{k\notin\emph{M}}{p_k 
\exp{\frac{i (E_k 
- \ensavg{E}) t}{\hbar}}}} - \frac{1}{\alpha^2} \\
&\geq& \abs{\sum_{k\notin\emph{M}}{p_k \exp{\frac{i \Delta E_k 
t}{\hbar}}}} - 
\frac{1}{\alpha^2} \\
&\geq& 
\sqrt{\left(\sum_{k\notin\emph{M}}{p_k \cos{\frac{\Delta E_k 
t}{\hbar}}}\right)^2 + 
\left(\sum_{k\notin\emph{M}}{p_k \sin{\frac{\Delta E_k 
t}{\hbar}}}\right)^2} - 
\frac{1}{\alpha^2} \\
&\geq& 
\sum_{k\notin\emph{M}}{p_k \cos{\frac{\Delta E_k 
t}{\hbar}}} -
\frac{1}{\alpha^2} \\
&\geq& \left(1-\frac{1}{\alpha^2}\right)\cos{\frac{\alpha \Delta E
t}{\hbar}}  - \frac{1}{\alpha^2}
\end{eqnarray}
where the last inequality holds provided that,
\begin{equation} \label{Equation::TCondition}
t < \frac{\hbar\pi}{2\alpha\Delta E}.
\end{equation}
QED
\end{proof}

\begin{proof}[Proof of Theorem~\ref{Theorem::7}]
Since $\ket{\psi_i^-}$ is an eigenstate of $H$, its evolution will 
correspond to multiplication by some overal phase 
$\exp{i\omega_{\Downarrow,i}^- t}$.  Then the remainder of the proof is 
trivial
\begin{eqnarray}
U\ket{\Downarrow,\psi_{\Downarrow,i}^-} 
&=& U \left(\ket{\psi_i^-}-\ket{\Uparrow,\psi_{\Uparrow,i}^-}\right) \\
&=& e^{i\omega_{\Downarrow,i}^-t}\ket{\Downarrow,\psi_{\Downarrow,i}^-} 
+\left(e^{i\omega_{\Downarrow,i}^- 
t}-U\right)\ket{\Uparrow,\psi_{\Uparrow,i}^-}.
\end{eqnarray}
QED
\end{proof}

\begin{proof}[Proof of Theorem~\ref{Theorem::8}]
Let $E_i^-$ be the eigenvalue corresponding to $\ket{\psi_i^-}$.  Let us 
consider the normalized trial state,
\begin{equation}
\ket{\phi_i} = \ket{\Uparrow,\psi_{\Downarrow,i}^-} + 
\ket{\Downarrow,\phi_{\Downarrow,i}}
\end{equation}
where
\begin{equation}
\ket{\phi_{\Downarrow,i}} = \frac{1}{-E_i^- - 
H_0}V\ket{\Uparrow,\psi_{\Downarrow,i}^-}.
\end{equation}
We select this trial state because it has the desirable property that 
$\Pi_\Uparrow\ket{\phi_i} = \ket{\Uparrow,\psi_{\Downarrow,i}^-}$.  Our 
selection of the spin-down component of $\ket{\phi}$ is designed to 
minimize the overall energy width of $\ket{\phi}$.  We then obtain,
\begin{eqnarray}
H\ket{\phi_i} &=& \left(H_0 + 
V\right)\left(\ket{\Uparrow,\psi_{\Downarrow,i}^-} 
+ \frac{1}{-E_i^- - H_0}V\ket{\Uparrow,\psi_{\Downarrow,i}^-}\right) \\
&=& H_0\ket{\Uparrow,\psi_{\Downarrow,i}^-} 
- E_i^-\frac{1}{-E_i^- -H_0}V\ket{\Uparrow,\psi_{\Downarrow,i}^-}
+ V\frac{1}{-E_i^- - H_0}V\ket{\Uparrow,\psi_{\Downarrow,i}^-} \\
&=& -R_{\pi} H_0\ket{\Downarrow,\psi_{\Downarrow,i}^-} 
- E_i^-\frac{1}{-E_i^- -H_0}V\ket{\Uparrow,\psi_{\Downarrow,i}^-}
+ V\frac{1}{-E_i^- - H_0}V\ket{\Uparrow,\psi_{\Downarrow,i}^-} \\
&=& -E_i^-\ket{\Uparrow,\psi_{\Downarrow,i}^-}+R_\pi V 
\frac{1}{E_i^- -H_0}V\ket{\Downarrow,\psi_{\Downarrow,i}^-} 
- E_i^-\frac{1}{-E_i^- -H_0}V\ket{\Uparrow,\psi_{\Downarrow,i}^-}
+ V\frac{1}{-E_i^- - H_0}V\ket{\Uparrow,\psi_{\Downarrow,i}^-} \\
&=& - E_i^-\ket{\phi_i}
+R_\pi V\frac{1}{E_i^- -H_0}V R_\pi\ket{\Uparrow,\psi_{\Downarrow,i}^-} 
+ V\frac{1}{-E_i^- - H_0}V\ket{\Uparrow,\psi_{\Downarrow,i}^-}
\end{eqnarray}
where we have used the fact that $R_\pi H_0 R_\pi = -H_0$.  If we define 
the operator
\begin{equation} \label{Equation::WDef}
W = \Pi_\Uparrow\left(R_\pi V\frac{1}{E_i^--H_0}V R_\pi + 
V\frac{1}{-E_i^- - H_0}V\right)\Pi_\Uparrow
\end{equation}
then we can evaluate the expectation value,
\begin{eqnarray}
\label{Equation::PhiHPhi}
\bra{\phi_i}H\ket{\phi_i} &=& -E_i^- + \bra{\phi_i}W\ket{\phi_i} \\
\label{Equation::PhiH2Phi}
\bra{\phi_i}H^2\ket{\phi_i} &=& (E_i^-)^2 - 
2E_i^-\bra{\phi_i}W\ket{\phi_i} + \bra{\phi_i}W^2\ket{\phi_i} \\
\ensavg{\left(\Delta H\right)^2} &=& \bra{\phi_i}W^2\ket{\phi_i} - 
\bra{\phi_i}W\ket{\phi_i}^2
\end{eqnarray}
These expectation values can be bounded using the L2 norm of $W$ (see 
\refeqn{L2NormDef} for a definition of the L2 matrix norm).  To 
obtain this norm, we write $W$ in the $\v{z}$ basis,
\begin{equation}
W = W_{diag} + W_{off-diag}.
\end{equation}
Here we define $W_{diag}$ as the diagonal portion of $W$ in the 
$\{\ket{\v{z}}\}$ basis,
\begin{equation}
W_{diag} =
\sum_{\v{z}}{\sum_{j,k}{\left(
\frac{\hbar^4 A_j^2}{4\left(E_i^- + E^\Downarrow_{0,\v{z}} + \hbar^2 A_j 
/ 2\right)}-\frac{\hbar^4 A_k^2}{4\left(E_i^- + E^\Downarrow_{0,\v{z}} + 
\hbar^2 A_k / 2\right)}\right)\ket{\v{z}}\bra{\v{z}}}}
\end{equation}
where $j$ sums over nuclei for which $z_j = -1$ and $k$ sums over 
nuclei such that $z_k = +1$.  $W_{off-diag}$ is the off-diagonal portion of 
$W$,
\begin{eqnarray} 
\label{Equation::Wnocancel}
W_{off-diag} &=&
\sum_{\v{z},\v{z'}}{\left(
\frac{\hbar^4 A_j A_k}{4\left(E_i^- + E^\Downarrow_{0,\v{z}} + \hbar^2 A_k 
/ 2\right)}-\frac{\hbar^4 A_j A_k}{4\left(E_i^- + E^\Downarrow_{0,\v{z}} + 
\hbar^2 A_j / 2\right)}\right)\ket{\v{z'}}\bra{\v{z}}}\\
\label{Equation::Wcancel}
&=& \sum_{\v{z},\v{z'}}{\left(
\frac{\hbar^4 A_j A_k(A_j - A_k)}{8\left(E_i^- + E^\Downarrow_{0,\v{z}} - 
\hbar^2 A_k / 2\right)\left(E_i^- + E^\Downarrow_{0,\v{z}} + 
\hbar^2 A_j / 2\right)}\right)\ket{\v{z'}}\bra{\v{z}}}
\end{eqnarray}
where $\v{z}$, and $\v{z'}$ are nuclear spin configurations which differ 
only at nuclear positions $j$ and $k$ such that that $z_j = -1, z_k = +1$  
and $z'_j = +1, z'_k = -1$.  We note that 
the two terms in \refeqn{Wnocancel} are nearly equal, leading to a 
significant cancellation in 
\refeqn{Wcancel}.  This cancellation is a result of the spin echo 
experiment, and will be discussed in \refsec{Conclusions}.

Because $\hbar^2 A_j, \hbar^2 A_k$ are 
negligible relative to $E^-_i + E^\Downarrow_{0,\v{z}}$, we can neglect 
these terms in the denominator.  We also note that $\abs{E_i^- + 
E^\Downarrow_{0,\v{z}}} \geq \abs{E^-_{max} + E^\Downarrow_{0,\max}}$.
Then using the inequality in \refeqn{Norm2X}, we can bound $\norm{W}_2$ as
\begin{equation} \label{Equation::NormW}
\norm{W}_2 \leq \frac{\hbar^4}{4\left(E^-_{max} + 
E^\Downarrow_{0,\max}\right)}A_{2,max}+
\frac{\hbar^6}{4\left(E^-_{max} 
+ 
E^\Downarrow_{0,\max}\right)^2}A_{2,max}A_{max}
\end{equation}
The second term in this equality is strictly less than the first term by a 
factor of $\epsilon$ (see \refeqs{Epsilon2Down}{Epsilon}).  So we can 
write the 
strict inequality,
\begin{equation} \label{Equation::W2L2}
\norm{W}_2 \leq 
\frac{\hbar^4}{4\left(E^-_{max} + 
E^\Downarrow_{0,\max}\right)}A_{2,max}\left(1+\epsilon\right)
\end{equation}
Hence it follows from \refeqs{PhiHPhi}{PhiH2Phi} and \refeqn{W2L2} that
\begin{equation}
\bra{\phi}H^2\ket{\phi} - \bra{\phi}H\ket{\phi}^2
\leq
\frac{\hbar^4}{4\left(E^-_{max} + 
E^\Downarrow_{0,\max}\right)}A_{2,max}\left(1+\epsilon\right)
\end{equation}
QED
\end{proof}

\begin{proof}[Proof of Theorem~\ref{Theorem::9}]
Since we know the energy width of the trial state $\ket{\phi_i}$, we can 
apply Theorem 6 to characterize its dynamics
\begin{equation} \label{Equation::UPhi}
U\ket{\phi} = e^{i\omega_{\Uparrow,i}^- 
t}\sqrt{1-\lambda_i^2}\ket{\phi}
+ \lambda_i\ket{r_{\Uparrow,i}^-}.
\end{equation}
In \refeqn{UPhi}, $\omega_{\Uparrow,i}^-$ is an overall phase 
corresponding to $\theta$ in Theorem 6.
$\lambda_i$ is the magnitude of the residual evolution, and is bounded by 
\refeqn{LambdaBound}.  In 
order to characterize the evolution of 
$\ket{\Uparrow,\psi_{\Downarrow,i}^-}$, we then note that
\begin{equation}
\ket{\Uparrow,\psi_{\Downarrow,i}^-} = \ket{\phi} - 
\ket{\Downarrow,\phi_{\Downarrow,i}}.
\end{equation}
Using similar reasoning as in the proof for Theorem 7, we obtain
\begin{eqnarray}
U\ket{\Uparrow,\psi_{\Downarrow,i}^-} &=& U\left(\ket{\phi_i} - 
\ket{\Downarrow,\phi_{\Downarrow,i}}\right) \\
&=& e^{i\omega_{\Uparrow,i}^- 
t}\sqrt{1-\lambda_i^2}\ket{\Uparrow,\psi_{\Downarrow,i}^-}
+ \lambda_i\ket{r_{\Uparrow,i}^-}
+ 
\left(\sqrt{1-\lambda_i^2}e^{i\omega_{\Uparrow,i}^-t}-U\right)\ket{\Downarrow,\phi_{\Downarrow,i}}
\end{eqnarray}
The final element in our proof is to demonstrate the three 
inequalities in \refeqto{InEq1}{InEq3}.  \refeqn{InEq1} follows 
directly 
from Theorem 6.  \refeqn{InEq2} follows from the fact that we can bound 
the magnitude of $\ket{\Uparrow,\phi_{\Uparrow_i}}$,
\begin{eqnarray}
\bkm{\Downarrow,\phi_{\Downarrow,i}}{\Downarrow,\phi_{\Downarrow,i}} &=&
\bra{\Uparrow,\psi_{\Downarrow,i}^-}V\frac{1}{\left(-E^-_i - 
H_0\right)^2}V\ket{\Uparrow,\psi_{\Downarrow,i}^-} \\
&\leq& \epsilon^2.
\end{eqnarray}
Finally, \refeqn{InEq3} follows from the consideration that (from Theorem 
6), $\bkm{r_{\Uparrow,i}^-}{\phi_i} = 0$.  Then,
\begin{eqnarray}
\abs{\bkm{r_{\Uparrow,i}^-}{\Uparrow,\psi_{\Downarrow,i}^-}} 
&=&  
\abs{\bkm{r_{\Uparrow,i}^-}{\phi_i}- 
\bkm{r_{\Uparrow,i}^-}{\Downarrow,\phi_{\Downarrow,i}^-}} \\
&=&  
\abs{\bkm{r_{\Uparrow,i}^-}{\Downarrow,\phi_{\Downarrow,i}^-}} \\
&\leq&  
\norm{\ket{r_{\Uparrow,i}^-}}\norm{\ket{\Downarrow,\phi_{\Downarrow,i}^-}} \\
&\leq& \epsilon.
\end{eqnarray}
QED
\end{proof}

\end{document}